\documentclass[journal]{IEEEtran}

\usepackage[linesnumbered,ruled,vlined]{algorithm2e}
\usepackage{multirow,booktabs,color,soul}
\usepackage{booktabs}
\usepackage{multirow}
\usepackage{subfigure}
\usepackage{epsfig}
\usepackage{amsmath}
\usepackage{amssymb}
\usepackage{tabularx}
\usepackage{makecell}
\usepackage{pdflscape}
\usepackage{array}

\ifCLASSINFOpdf
\else
\fi

\begin{document}

	\title{Logistics in the Sky: A Two-phase Optimization Approach for the Drone Package Pickup and Delivery System}

	\author{Fangyu Hong, Guohua Wu,~\IEEEmembership{Member,~IEEE}, Qizhang Luo, Huan Liu, Xiaoping Fang, Witold Pedrycz,~\IEEEmembership{Life Fellow,~IEEE} 
		
	\thanks{This work was supported by the National Natural Science Foundation of China under Grant 62073341, Natural Science Fund for Distinguished Young Scholars of Hunan Province under Grant 2019JJ20026, and the China Scholarship Council (No. 202006370285).}
		
	\thanks{Fangyu Hong, Guohua Wu (\textit{the corresponding author}), Qizhang Luo, Huan Liu and Xiaoping Fang are with the School of Traffic and Transportation Engineering, Central South University, Changsha 410075, China. E-mail: fangyuhong@csu.edu.cn; guohuawu@csu.edu.cn; qz\_luo@csu.edu.cn; liuhuan1095@csu.edu.cn; fangxp@csu.edu.cn;  }

	\thanks{Witold Pedrycz is with the Department of Electrical and Computer Engineering, University of Alberta, Edmonton, AB T6G 2V4, Canada, with the Department of Electrical and Computer Engineering, Faculty of Engineering, King Abdulaziz University, Jeddah 21589, Saudi Arabia, and also with the Systems Research Institute, Polish Academy of Sciences, Warsaw 01447, Poland. E-mail: wpedrycz@ualberta.ca.}
		
	}
	

	\markboth{}%
	{Fangyu \MakeLowercase{\textit{et al.}}: Bare Demo of IEEEtran.cls for IEEE Journals}

	\maketitle

	\begin{abstract}

	The application of drones in the last-mile distribution is a research hotspot in recent years. Different from the previous urban distribution mode that depends on trucks, this paper proposes a novel package pick-up and delivery mode and system in which multiple drones collaborate with automatic devices. The proposed mode uses free areas on the top of residential buildings to set automatic devices as delivery and pick-up points of packages, and employs drones to transport packages between buildings and depots. Integrated scheduling problem of package drop-pickup considering \textit{m}-drone, \textit{m}-depot, \textit{m}-customer is crucial for the system. We propose a simulated-annealing-based two-phase optimization approach (SATO) to solve this problem. In the first phase, tasks are allocated to depots for serving, such that the initial problem is decomposed into multiple single depot scheduling problems with \textit{m}-drone. In the second phase, considering the drone capability constraints and task demand constraints, we generate the route planning scheme for drones in each depot. Concurrently, an improved variable neighborhood descent algorithm (IVND) is designed in the first phase to reallocate tasks, and a local search algorithm (LS) are proposed to search the high-quality solution in the second phase. Finally, extensive experiments and comparative studies are conducted to test the effectiveness of the proposed approach. Experiments indicate that the proposed SATO-IVND can reduce the cost by more than 14\% in a reasonable time compared with several other peer algorithms.

	\end{abstract}

	
	\IEEEpeerreviewmaketitle

	\section{Introduction}

	\IEEEPARstart{T}{HE} last link of logistics transportation is last-mile package delivery \cite{2014Hierarchical}. In this part, logistics enterprises need to communicate with customers directly and deliver package to them. The rapid development of e-commerce has driven the development of express industry, and also promoted continuous changes of package delivery mode. As a new delivery tool, drones have been increasingly considered in the last-mile delivery because of its flexibility, low cost and high reliability. Since the traffic congestion occurs frequently in the urban environment, classical delivery tools may be inefficient in rush hours. In contrast, drones have obvious advantages in urban package delivery. Firstly, drones are not restricted by road conditions and can deliver package without drivers, so that carrying packages by drones can be done all day. In addition, drones can reach areas that are inaccessible to trucks, such as residential areas.

	The common settings of the last-mile delivery include couriers, delivery stations, intelligent express cabinets, etc. With the development of the e-commerce business and the expansion of cities, the shortcomings of these models are becoming more and more obvious. Couriers delivers or picks up packages door to door. However, delivery or pickup by couriers is less efficient and has high human cost. Delivery by delivery stations or intelligent express cabinets is more efficient to deal with packages. Customers can go to the sites to pick up or deliver their packages. However, the location of setting delivery stations and intelligent express cabinets are usually restricted by limited space in urban. Additionally, the customers who live far from the delivery station and intelligent express cabinet need to spend much time and resources to pick up packages.

	In order to deal with the above problems in the last-mile delivery, many scholars began to design new distribution modes in which the application of drone has attracted much attention. Amazon started trying to deliver small packages by drones in 2013 \cite{RN02}. Subsequently, several other companies have begun similar research. DHL is trying to deliver medicine by drones to people who are living in remote areas \cite{RN03}. Many Chinese enterprises, such as Meituan, Jingdong, Shunfeng , are also committed to research on providing just-in-time delivery services by drones \cite{RN04,RN06}.

	Compared with the traditional delivery modes, deliverying with drones can effectively ease the pressure of the manual delivery. As drones are flexible and easy to operate, delivery with drones is almost unlimited by traffic conditions. Obviously, this mode of delivery can deliver packages successfully and potentially bring better service quality for the last-mile delivery, even in a crowded city. Nevertheless, due to the current power restriction in long distance delivery, single drone delivery is limited to some constraints, such as the flying range constraint \cite{2020Joint}. and load capacity constraint \cite{2021A}. Therefore, the limited battery capacity of drones is the current challenge for delivering many packages in a large area. To address this issue, scholars have developed two common settings for drone delivery.

	The first one is that the drone is only used to deliver packages. Dorling \textit{et al.}  \cite{2017Vehicle} proposed that optimizing battery weight and payload weight can reduce the cost for drone delivery. Song \textit{et al.}  \cite{2018Persistent} proposed persistent UAV delivery schedules that drones are allowed to share multiple stations to replenish their consumables. This mode may have security risk in urban environment when drones with delivery packages take off and land from customers.

	The other one is the combined truck-drone delivery. In 2015, Murray and Chu \cite{murray2015flying} proposed a delivery mode which used a truck and a drone working cooperatively for delivering packages. Phan \textit{et al.} \cite{tu2018traveling} proposed Traveling Salesman Problem with Multiple Drones (TSP-\textit{m}D), and employed trucks and drones to deliver packages in parallel. However, there are still many challenges in this mode. For example, it is difficult to arrange a parking stop for trucks and security risk may exist when drones takeoff and landing from trucks or customers.
	
	However, different from the above modes, we propose a novel last-mile delivery mode and system. We named this system as the Drone Package Pickup and Delivery System, which will make good use of the free area at the top of the residential building. In our system, we set the automatic devices at the top of the building to deposit packages and use drones to deliver packages between buildings and depots. When the package is transported to the right place, the automatic device can identify the unique barcode of the package and deposit the package. In addition, the device will inform customers to pick up their packages. In order to make it more convenient for customers to send packages, customers can make an appointment of delivery service at home or anywhere they like. After customers put their packages on the automatic device, the system will arrange drones to pick up and transport the packages. On the one hand, the whole process is carried out on the roof, so it is not limited by the conditions of urban and it is more security during for drones to takeoff and land. On the other hand, the whole delivery process is almost completed by drones and automation devices. Therefore, it can reduce the human cost of delivery, improve customers' satisfaction, and promote intelligent distribution in last-mile. The whole system is efficient and robust.

	The integrated scheduling with \textit{m}-drone, \textit{m}-customer and \textit{m}-depot is crucial for the Drone Package Pickup and Delivery System. This problem has been proved to be an NP-hard problem \cite{shima2006multiple}. Exact algorithms can obtain the optimal solution of the problem at the cost of high computational complexity and much running time \cite{2018Integrated}. In contrast, heuristic algorithms \cite{2018A,2019An} and metaheuristic algorithms \cite{ancele2021toward} are popular and promising alternatives for solving large-scale problem. Currently, package delivery with drones is taken into account in relevant research, while package pick-up is scarcely mentioned. In addition, for the integrated scheduling problem considering drop-pickup, few algorithms can find high-quality solution in a reasonable time.

	Hence, we designed a simulated-annealing-based two-phase optimization approach (SATO) to solve this problem. This approach divides the initial problem into a task allocation phase of \textit{m}-depot and a route planning phase of each single depot. In the task allocation phase, we generate a task allocation scheme for each depot, such that the original problem is decomposed into route planning problems. In the route planning phase, we plan pick-up and delivery route of drones based on the task allocation scheme. The route planning of drones must meet the drone capacity constraints and task demand constraints. These two phases are executed iteratively and interactively until the predefined stopping criteria are satisfied. Under the framework, in the task allocation phase, we employ $k$-means algorithm to generate an initial task allocation scheme, and propose an improved variable neighborhood descent algorithm (IVND) to reallocate tasks. In the route planning phase, the effect of payload on drone fly range is considered and heuristic rules are used to construct route planning for each single depot. Besides, a local search algorithm is designed (LS) for exploit better route planning. Consequently, an iterative two-phase optimization method, named SATO-IVND for short, is proposed for the integrated scheduling problem with \textit{m}-drone, \textit{m}-customer and \textit{m}-depot. The proposed two-phase approach (SATO) can effectively reduce the complexity of the original problem, while IVND and LS can explore and exploit satisfactory solution.

	The main contributions of this paper are highlighted as follows:

	\begin{itemize}
		
		\item We propose a novel drone package pickup and delivery mode and system to realize agile and efficient last-mile delivery. As the limitations of space in urban environment and security risk during takeoff and landing of drones, in the system, automatic devices are placed in the free area on the top of the residential building. We use automatic devices as the delivery and pick-up point of packages and use drones to transport packages between buildings and depots.
		
		\item To address the scheduling and routing problem in the proposed system, we develop a simulated-annealing-based two-phase optimization approach (SATO). In the first phase, we generate a task allocation scheme for each depot. In the second phase, route planning of drones is generated for every depot according to the task allocation scheme generated in the first phase. These two phases are executed iteratively and interactively until the predefined stopping criteria are satisfied. This approach can reduce the computational complexity and solve problems efficiently.
		
		\item We propose an improved variable neighborhood descent (IVND) to assist the task allocation in the first phase and a local search algorithm (LS) to design the route planning in the second phase. IVND algorithm is designed for generating task allocation scheme considering the drone capability constraints and task demand constraints. LS is used to search a satisfactory scheme based on the current best scheme generated by IVND. It can find the high-quality scheduling schemes effectively.
		\item We conduct extensive experiments to validate the efficiency of the proposed SATO-IVND algorithm. The experimental results show that SATO-IVND is superior to the other six heuristics and metaheuristics with respect to solution quality and computing overhead, especially for large-scale problems. In addition, under a realistic scenario with 80 tasks, the proposed SATO-IVND generates a high-quality scheduling scheme, demonstrating its effectiveness and practicability. 
		
	\end{itemize}

	The rest of this paper is organized as follows: Section II reviews the related work. Section III constructs the model of the drone package delivery and pickup. Section IV describes the two-phase optimization approach we proposed. Section V covers the computational experiments result. Section VI provides conclusions and future study trend.

	\section{Related Studies}

	With the maturity of technology, drones are widely used in various traffic fields, such as package delivery, traffic data acquisition and traffic surveillance \cite{das2020synchronized}. Among them, package delivery with drones is a research hotspot in recent years. Also, with the increase of online shopping, the reverse logistics caused by returning and changing items are lack of scientific solutions. Thus, our research mainly focuses on the application of drones in last-mile package delivery and pickup.

	To the best of our knowledge, the research on the application of drones in package delivery mostly focuses on two modes: the independent use of drones or ``truck \& drone'' \cite{2019Review}. For example, Mathew \textit{et al.} \cite{mathew2015planning} established a cooperative package delivery mode with a truck and a drone, in which customers can be served by the drone. Also, as an auxiliary, the truck can provide packages transportation, charging and other services for the drone. Then the work was later extended by Karak and Abdelghany to consider delivering and picking up package by drone and truck \cite{karak2019hybrid}. Wang \textit{et al.} \cite{wang2019expressway} addressed a setting that customers can be served by trucks or drones. Besides, in their design, trucks can also serve as a depot and a landing platform for drones. Different from previous studies, Dayarian \textit{et al.} \cite{2017Same} introduced a new package delivery mode that trucks are used to deliver packages for customers and are resupplied by drones.

	Integrated scheduling of \textit{m}-depot, \textit{m}-customer, \textit{m}-drone is a kind of combinatorial optimization problem satisfying some constraints such as drop-pickup, the drone capacity constraints and task demand constraints. Like solving common integrated scheduling problems of \textit{m}-drones, task allocation and route planning are two main parts to deal with the initial problem \cite{chen2021clustering}. Task allocation will determine which depot the tasks will be assigned to, while route planning will determine the order of task access. Previous work tried to solve integrated scheduling problems of \textit{m}-drone as a whole, which makes it hard to generate high-quality scheduling schemes in a reasonable time. To solve the issue, some studies focused on the innovation of scheduling framework when addressing complex scheduling problems. For example, Deng \textit{et al.} \cite{deng2020two} proposed a two-phase coordinated planning approach for heterogeneous Earth-observation resources, which included an area target decomposition phase and a task allocation phase. Liu \textit{et al.} \cite{2020Iterative} proposed a divide and conquer framework to solve multi-drone task scheduling problems, which included a task allocation phase and a single drone scheduling phase.

	The search algorithms for task allocation mainly include heuristic algorithm and metaheuristic algorithm like clustering algorithms \cite{ma2021unsupervised},distributed algorithm based on market mechanism  \cite{lee2014resource}, randomized greedy-algorithm \cite{ha2018min}, genetic algorithm (GA) \cite{moore2007distributed} etc. At present, distributed algorithm based on market mechanism is widely used in task allocation. For example, Lee \textit{et al.} \cite{lee2014resource} proposed a decentralized auction algorithm for task allocation. In each round of auction, the robots bid for its most ideal task, and the decision system would determine the ownership of the task. This algorithm exhibits high communication requires and the communication cost may be a little high. Metaheuristic algorithms, such as GA, need continuous iterative calculation to deal with the task allocation problem. Randomized greedy-algorithm and clustering algorithms are often invoked in the initial task allocation phase. The key of randomized greedy-algorithms is the choice of greedy strategy, while Clustering algorithm can generate task allocation schemes considering the geographic locations of tasks.

	The search algorithms for route planning mainly include exact algorithms \cite{roberti2021exact}, heuristic and metaheuristic algorithms \cite{han2020metaheuristic}. Exact algorithms are capable of finding the optimal solution of small-scale problems. However, exact algorithms may not be suitable for solving large-scale problems because of its high computational complexity \cite{hong2018range}. Heuristic algorithms and metaheuristic algorithms could be more suitable for finding satisfactory solutions of large-scale problems. Metaheuristic algorithms can be applied to different problems, but the optimization effect could be unstable without sophisticated design. Heuristic algorithms are generally oriented specific problems, but they depend on specific heuristic rules. Extensive works have been done to improve the efficiency of the algorithms. Liu \textit{et al.} \cite{2020Iterative}applied tabu list to simulated annealing algorithm (SA), which prohibited the circulation or repetition of solutions in a short time, and experiments demonstrated that it can effectively reduce the consumption of solving time. Peng \textit{et al.} \cite{peng2019hybrid} proposed a hybrid genetic algorithm to resolve traveling salesman problem with multiple drones. Gao \textit{et al.} \cite{gao2020learn} used deep reinforcement learning to train the destroy and repair operators of large neighborhood search algorithm (LNS), and its good performance is proved by experiments. Kitjacharoenchai \textit{et al.} \cite{kitjacharoenchai2020two} proposed a LNS to find the promising route of trucks and drones. Li \textit{et al.} \cite{li2020two} proposed a LNS algorithm for scheduling of trucks and drones. Ferrandez \textit{et al.} \cite{ferrandez2016optimization}  adopted $k$-means and genetic algorithm to optimize the route planning of truck and drone.

	Based on the analyses of the above existing research progress, it can be found that:

	(1) Most of the literatures focus on the application of drones to deliver packages without considering picking up packages at the same time.

	(2) Previous works try to solve integrated scheduling problem of \textit{m}-drone, \textit{m}-customer and \textit{m}-depot as a whole, which result in classical heuristic algorithm and metaheuristic algorithm hardly can generate satisfactory solutions time efficiently.

	(3) Combined truck-drone delivery is more suitable in rural areas, while in urban environment the mode may be difficult to arrange stops for trucks and have security risk during drones' takeoff and landing.

	In brief, we propose a novel package pick-up and delivery mode and system, then we construct a mixed integer linear programming (MILP) model for integrated scheduling problem of \textit{m}-drone, \textit{m}-customer and \textit{m}-depot considering drop-pickup. After that, we propose a two-phase optimization framework to help solve the original problem. Finally, we design SATO-IVND algorithm, which shows good performance in dealing with large-scale task allocation and route planning problem.

	\section{Model Of The Drone Package Delivery And Pickup}

	\subsection{Problem Assumptions}

	\begin{figure*}[htb]
		\begin{center}
			\subfigure{\psfig{file=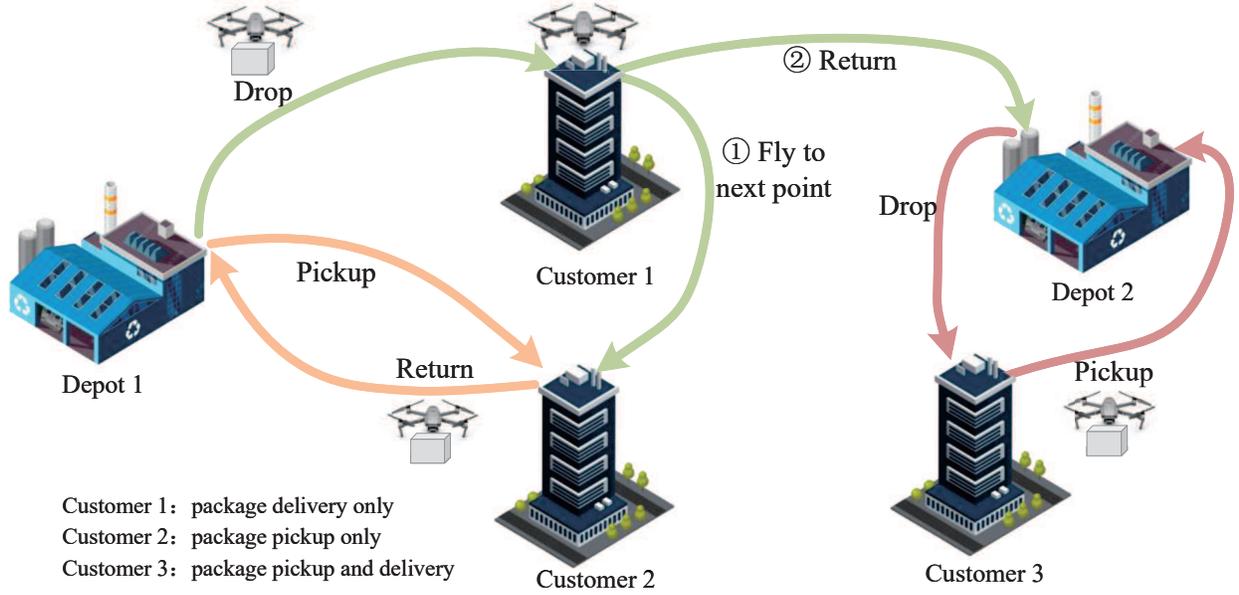, angle=0, width=6.6in}}
		\end{center}\vspace{-5mm}
		\caption{The Drone Package Pickup and Delivery Mode and System.} \label{Fig:Fig1-hong}
	\end{figure*}

	The Drone Package Pickup and Delivery System for urban last-mile distribution in this study consists of \textit{m}-drone, \textit{m}-depot, \textit{m}-customer. We suppose that:
	(1) there are multiple drone stations in the city; 
	(2) each station has multiple drones;
	(3) each drone station can also be regarded as a depot.

	The roof of urban residential building is used as the delivery/pickup point of drones. Each delivery/pickup point can be regarded as a customer. Additionally, each customer may contain many tasks of package delivery and pickup. Also, each customer can be visited many times. Automatic devices are placed on the roof to store, load and unload packages. The drone can load or unload packages on the roof and then go to the next location. In addition, the drone can replace its battery on the top of the residential building automatically, and the replaced battery can be charged on the roof, so as to improve the transportation efficiency and enlarge drone flying range.

	As shown in Fig.~\ref{Fig:Fig1-hong}, the system has multiple customers, multiple depots and multiple drones. The green line indicates that drone can deliver packages from the depot to the customer and drop packages, then the drone can return to depot or fly to next customer. The red line indicates that drone can deliver package from the depot to customer and drop package; if customer need package pickup service, the drone will pick up package from the customer and return to depot. The orange line indicates that drone can fly to customer for picking up package and return to depot. The problem can be regarded as integrated scheduling problem of \textit{m}-depot, \textit{m}-drone, \textit{m}-customer considering drop-pickup.

	During the drone delivery, drones can encounter some obstacles in the urban environment. For obstacle avoidance of drone, we can refer to the method of Liu \textit{et al.} \cite{liu2020autonomous} , which proposed an autonomous path planning method. When drone encounters obstacle, the method generates two paths between two points based on tangent intersection and target guidance strategy. Then one of the paths is selected according to the heuristic rules (distance, obstacle avoidance conditions, etc.).

	The assumptions of this paper are as follows:
	
	(1) The packages transported by drones are packed in special boxes of uniform size.
	
	(2) The charging time of drones is not considered.
	
	(3) Drone will fly directly to the customer without detour.
	
	(4) Drone can only carry one package each time.
	
	(5) Drones fly with constant speed without considering the energy consumption of drone take-off and landing.

	\subsection{Parameters}

	\definecolor{hl}{rgb}{0.75,0.75,0.75}
	\sethlcolor{hl}
	\begin{table}[htp]
		\caption{Description of main notations} \label{Table:table1} \vspace{-3mm}
		\footnotesize
		\renewcommand{\arraystretch}{1.5}
		\begin{center}
			\begin{tabular}{ll}
				\toprule
				Notations & Description \\ \midrule
				$B$        & A set of depots, $B={1,2,\cdots,m}$ \\ 
				$U$        &A set of drones, $U={1,2,\cdots,k}$ \\ 
				$C$        & A set of tasks, $C={1,2,\cdots,c}$ \\  
				$k$        & Index of drone \\ 
				$n$        & Index of package \\ 
				$T_i$        & Task type of task $i$ \\ 
				$C_n$        & The weight of package $n$ \\ 
				$C_{max}$        & Maximum capacity of drone \\ 
				$h$        & Maximum flying range of drone \\ 
				$d_{i,j}$        & The distance between point $i,j$ \\
				\bottomrule
			\end{tabular}
		\end{center}
	\end{table}

	Basic element description: let $m$ denote the number of depots. Let $c$ denote the number of tasks. In order to distinguish three different task types including package pick-up only, package delivery only, package pick-up and delivery simultaneously, we define three sets:
	\textit{DROP}, \textit{PICKUP}, \textit{PICK-DROP}.
	\textit{DROP} denotes the set of tasks which need package delivery service only. \textit{PICKUP} denotes the set of tasks which need package pick-up service only. \textit{PICK-DROP} denotes the set of tasks which need package pick-up and delivery service concurrently. In order to distinguish two different operations of package pickup or delivery in the same task, we factorize each task into two virtual tasks. Two virtual task sets $C^{pick}=\{1,2,\cdots,c\}$ and $C^{drop}=\{c+1,c+2,\cdots,2c\}$ are created for the initial task set $C$. For example, the original task $i\in C$ corresponds to task $i\in C^{pick}$ and task $i\in C^{drop}$. The main notation used in this paper are listed in Table I.

	We define a binary variable $x_{i,j}^k$, which denotes whether the drone $k$  flies from task $i$ to task $j$. If the drone $k$ flies from task $i$ to task $j$, $x_{i,j}^k=1$. Otherwise, $x_{i,j}^k=0$.
	
	We define a binary variable $u_k$ to denote whether the drone $k$ start or not. If the drone $k$ is used for completing tasks, $u_k=1$. Otherwise, $u_k=0$.

	We define an integer variable $T_i$ to denote the task type of task $i$. If task $i$ need package delivery service, then we have $T_i=-1$; If task $i$ need package pickup service, we have $T_i=1$; $T_i=0$ means that task $i$ have no need of package delivery/ pickup.

	We define a variable $h_n^k$ to donate the flying range of drone $k$ when it carries package $n$. The flight time of drone decreases linearly with the increase of payload \cite{RN43}.There is a linear correlation between flight time and flying range when drone is in uniform speed. Therefore, there is a linear correlation between the flying range and payload of drone. $h_n^k$ can be formulated as follows:

	\begin{equation}\label{eq1}
	\begin{array}{cc}
	h_n^k=\frac{h}{\beta}
	\end{array}
	\end{equation}
	where $\beta$ is the payload penalty factor of drone. The flying range and energy consumption of drones are significantly affected by payload \cite{jeong2019truck}.When the drone power is constant, the flying range of drone decreases linearly with the increase of payload. Refer to the weight function of flight time based on the number of packages loaded in literature \cite{2018Persistent}, $\beta$ is given as follows:
	
	\begin{equation}\label{eq2}
	\begin{array}{cc}
	\beta(C_n)=\frac{\beta_{max}-1}{C_{max}}C_n+1
	\end{array}
	\end{equation}
	where $\beta_{max}$ is the maximum of $\beta$. When the drone is empty, $\beta=1$. Meanwhile,$\beta=\beta_{max}$ when the drone is fully loaded.
	
	\subsection{Model}

	The original problem of this paper is an integrated scheduling of \textit{m}-depot, \textit{m}-customer, \textit{m}-drone constrained by drop-pickup, flying range and payload, etc. It can be seen as an \textit{m}-drone parallel scheduling traveling salesman problem (\textit{m}D-PSTSP). In order to find a high-quality solution, we build a model to minimize the total cost of drones, mainly including the flight costs of drones and the number of drone launch sorties \cite{yoon2018traveling}.

	The model of the drone package delivery and pick-up can be formulated as follows:

	\begin{equation}\label{eq3}
	\begin{array}{cc}
	\mbox{Minimize} f=\alpha\underset{i \in B\cup C}{\sum}\underset{j \in B\cup C,j\neq i}{\sum}\underset{k \in U}{\sum}d_{i,j}\cdot x_{i,j}^k\\
	+\rho\underset{k \in U}{\sum}u_k\\
	\end{array}
	\end{equation}
	
	\textit{C}1:
	\begin{equation}\label{eq4}
	\begin{array}{cc}
	x_{i,j}^k \in \{0,1\},\forall k \in U,i,j \in {B\cup C}
	\end{array}
	\end{equation}
	
	\textit{C}2:
	\begin{equation}\label{eq5}
	\begin{array}{cc}
	T_i \in \{0,1\},\forall i \in C^{pick}
	\end{array}
	\end{equation}
	\begin{equation}\label{eq6}
	\begin{array}{cc}
	T_i \in \{0,-1\},\forall i \in C^{drop}
	\end{array}
	\end{equation}
	\begin{equation}\label{eq7}
	\begin{array}{cc}
	T_i+T_{i+c}=-1,\forall i \in {DROP}
	\end{array}
	\end{equation}
	\begin{equation}\label{eq8}
	\begin{array}{cc}
	T_i+T_{i+c}=1,\forall i \in {PICKUP}
	\end{array}
	\end{equation}
	\begin{equation}\label{eq9}
	\begin{array}{cc}
	T_i+T_{i+c}=0,\forall i \in {PICK-DROP},T_i\neq T_{i+c}
	\end{array}
	\end{equation}
	
	\textit{C}3:
	\begin{equation}\label{eq10}
	\begin{array}{cc}
	\underset{k \in U}{\sum}\underset{b \in B}{\sum}x_{b,i}^k=1,\forall i \in C
	\end{array}
	\end{equation}
	\begin{equation}\label{eq11}
	\begin{array}{cc}
	\underset{k \in U}{\sum}\underset{b \in B}{\sum}x_{i,b}^k=1,\forall i \in C
	\end{array}
	\end{equation}
	
	\textit{C}4:
	\begin{equation}\label{eq12}
	\begin{array}{cc}
	\underset{j \in {B\cup PICKUP}}{\sum}x_{i,j}^k=1,\forall i \in DROP,k \in U
	\end{array}
	\end{equation}
	\begin{equation}\label{eq13}
	\begin{array}{cc}
	\underset{b \in B}{\sum}x_{i,b}^k=1,\forall i \in PICKUP,k \in U
	\end{array}
	\end{equation}
	\begin{equation}\label{eq14}
	\begin{array}{cc}
	\underset{k \in U}{\sum}\underset{b \in B}{\sum}x_{i,b}^k=1,\forall i \in PICK-DROP,k \in U
	\end{array}
	\end{equation}
	
	\textit{C}5:
	\begin{equation}\label{eq15}
	\begin{array}{cc}
	d_{i,j}\le h_n^k,\forall i,j \in {B\cup C},k \in U
	\end{array}
	\end{equation}

	\textit{C}6:
	\begin{equation}\label{eq16}
	\begin{array}{cc}
	c_n\le c_{max}
	\end{array}
	\end{equation}
	In our model, the objective function $f$(Eq. (3)) aims to minimize the total cost of drones. The first part of $f$ is to minimize the travel cost of drones, as the longer the drone flies, the more energy it consumes; the second part of $f$ is to minimize the number of drone launch sorties. Besides, $\alpha$ and $\rho$ ($\alpha,\rho \in [0,1]$) are the weight coefficients of the two parts respectively \cite{han2020metaheuristic}.

	The decision variables of the proposed model are prescribed in constraint \textit{C}1. Constraint \textit{C}2 prescribes the type of task $i$. Constraint \textit{C}3 means that the drone must start from the depot and finally return to the nearest depot after completing all its scheduled tasks. Constraint \textit{C}4 prescribes the constraints about tasks. Eq. (12) means that after drone completes the delivery task, it can choose to go to the next point with the demand of package pick-up or return to the nearest depot; Eq. (13) means that the drone shall return to the nearest depot after completing the package pick-up task, since the drone can only carry one package at a time; Eq. (14) means that the task which has demands of package delivery and pick-up simultaneously shall be serviced by one drone. Then, constraint \textit{C}5 represents flying range constraint, which means the distance between task $i$ and $j$ should not exceed the maximum fly range of the drone. Constraint \textit{C}6 represents drone loading constraint, the weight of package loaded by drone cannot exceed the maximum load of drone.

	\section{Two-phase Optimization Approach}
	
	The integrated scheduling of \textit{m}-depot, \textit{m}-customer, \textit{m}-drone constrained by drop-pickup, flying range and payload etc., is a complex combinatorial optimization problem. In this section, we will design a two-phase optimization method to solve this kind of problem, thus improving the efficiency of finding a high-quality solution.
	
	\subsection{Algorithm Framework}

\begin{figure}[!htb]
		\begin{center}
			\subfigure{\psfig{file=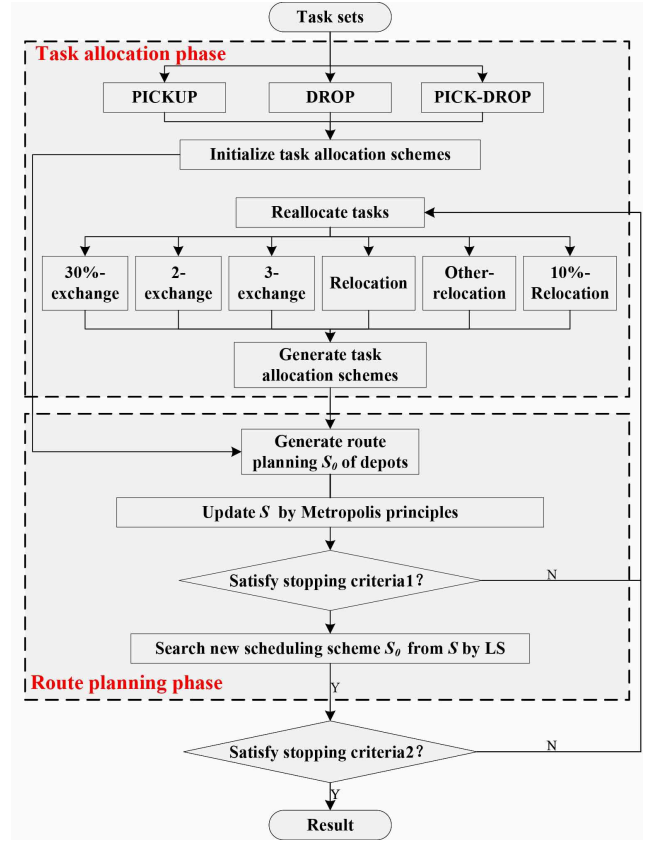, angle=0, width=3.3in}}
		\end{center}\vspace{-5mm}
		\caption{The framework of SATO-IVND.} \label{Fig:Fig2-hong}
	\end{figure}

	With the increasing number of tasks, the complexity of combinatorial optimization problem increases sharply. Traditional heuristic algorithms and metaheuristic algorithms are difficult to find satisfactory solution in a reasonable time. Thus, we design a SATO-IVND algorithm to resolve the original problem. The simulated-annealing-based two-phase optimization approach (SATO) decomposes the original integrated scheduling problem into two phases: a task allocation phase of \textit{m}-depot and a route planning phase of each single depot. In the first phase, the original \textit{m}-depot integrated scheduling problem is transformed into multiple single depot route planning problem; in the second phase, multiple route planning schemes for each single depot are found to complete all tasks. The two stages are iteratively and interactively performed in order to find a high-quality solution. The framework of two-phase optimization approach SATO-IVND is shown in Fig.~\ref{Fig:Fig2-hong}. The pseudocode of the SATO-IVND is shown as Algorithm 1.
	
	\begin{algorithm}
		\caption{\textit{SATO-IVND}}\label{PC:al1}
		\KwIn{task information;\mbox{ }distance between each pair of locations;\mbox{ }iteration gap $L$}
		\KwOut{route planning of all depots $S$;}
		Initialize a task allocation scheme $G\gets{G_1,G_2,\cdots,G_m }$ for all depots by $k$-means algorithm\;
		Generate a route planning scheme $S_0$ based on $G$ by ERPA\;
		Calculate initial the objective function value $f(S_0)$\;
		Let $S\gets S_0,f(S)\gets f(S_0)$\;
		
		\While{stopping criteria 1 is not satisfied}
		{
			\While{stopping criteria 2 is not satisfied}
			{
				Generate a new allocation scheme $G'$ from $S$ by IVND\;
				Generate a new route planning scheme $S_0$ based on $G'$\;
				Calculate the objective function value $f(S_0)$\;
				\If{the conditions of Metropolis for accepting the new scheme are met}
				{
					$S\gets S_0,f(S)\gets f(S_0)$\;
				}
			}
			Generate a new scheduling scheme $S_0$ from $S$ by LS\;
			Calculate the objective function value $f(S_0)$\;
			\If{$f(S_0)$ is superior to $f(S)$}
			{
				$S\gets S_0,f(S)\gets f(S_0)$\;
			}
		}
	\end{algorithm}

	In task allocation phase, we divide all tasks into three types: package pickup only, package delivery only and package pickup \& drop concurrently. After classification of all tasks, the \textit{k}-means algorithm is used to generate an initial task allocation scheme considering the geographical location of tasks. Then, the integrated scheduling problem of multiple depots is transformed into single depot scheduling problem. According to the scheduling results, we design an improved variable neighborhood descent algorithm (IVND) to adjust the allocation scheme. Simultaneously, six neighborhood transformation operators (i.e., 2-exchange, 3-exchange, 30\%-exchange, relocation, other-relocation and 10\%-relocation) are designed in IVND to adjust the allocation scheme. 2-exchange, 3-exchange and 30\%-exchange are used to adjust the allocation scheme in a depot. Two or more drones in a depot can exchange tasks assigned to them. Relocation, other-relocation and 10\%-relocation are used to adjust the task allocation scheme between two depots. The result generated by IVND will be accepted as the new allocation scheme if it satisfies the condition of Metropolis principles. The Metropolis principles \cite{2017Vehicle} are as follows:
	
	\begin{equation}\label{eq17}
	P =\left\{
	\begin{array}{cc}
	\mbox{ }  1,  \mbox{ } {\rm if} \mbox{ } df < 0,\\
	\mbox{ } exp(-(\frac{df}{t}))  \mbox{ } , \mbox{ } \mbox{otherwise}\\
	\end{array} \right.
	\end{equation}
	Where $df=f(S_0)-f(S)$, $f(S_0)$ is the objective function value of new solution and $f(S)$ is the objective function value of the previous solution. Besides, $P$ is the probability of accept a new solution. If $df<0$ then we accept the new solution with probability 1; otherwise, we accept the new solution with probability $exp(-(\frac{df}{t}))$.

	Then we need to generate ordered route planning scheme in the second phase. In the second phase, based on the result of task allocation derived in the first phase, some rules are used to generate the initial route planning for each single depot considering all constrains. Additionally, the LS algorithm is designed to refine the result of route planning. It is worth noting that the elitist mechanism would decide whether to accept the new solution generated by LS. The whole scheduling scheme are formed from the scheduling scheme of each single depot.

	The two phases iterate until the termination condition of the algorithm is satisfied. There are two loops in the algorithm: the inner cycle uses IVND algorithm to generate new allocation scheme satisfying various constraints. The inner cycle is executed until the maximum number of iterations is satisfied. The external cycle is to find the promising scheduling scheme based on the result generated by the inner cycle. The external cycle is performed until the current temperature reaches the lowest temperature of the SA algorithm.

	\subsection{Task Allocation}
	
	\subsubsection{Initial task allocation}
	
	\textit{k}-means algorithm is one of the commonly used algorithms in the field of unsupervised learning. Especially, \textit{k}-mean algorithm aims to divide the data points into multiple clusters, so that the sum of squares of the distance between the sample points in each cluster and the cluster center is the smallest. To solve the original problem, considering geographical position of each task, the \textit{k}-means algorithm is used to assign tasks to different depots according the distance between depots and tasks. Then each single depot obtains a task allocation scheme containing many tasks. In this way, the original \textit{m}-depot scheduling problem is partitioned into multiple single depot scheduling problem. Specifically, the result of initial task allocation is disordered and cannot be directly handed over to the drones for execution.

	\subsubsection{Neighborhood transformation operators in IVND}

\begin{figure}[!htb]
		\begin{center}
			\subfigure{\psfig{file=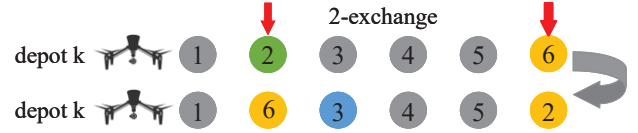, angle=0, width=3.3in}}
		\end{center}\vspace{-5mm}
		\caption{2-exchange.} \label{Fig:Fig3-hong}
	\end{figure}

	Based on the allocation scheme, we propose a IVND algorithm to find the satisfactory solution. Six neighborhood transformation operators are designed in IVND to reallocate the task. The operators include 2-exchange, 3-exchange, 30\%-exchange, relocation, other-relocation, 10\%-relocation. The description of six operators is given as follows:

	a. 2-exchange: exchange two tasks assigned the same depot. More specifically, the operation of 2-exchange is as follows: firstly, we choose depot $k$ from all depots randomly. Then, two tasks $i$ and $j$ allocated to depot $k$ are selected randomly ($i,j \in PICKUP$). Exchange the positions of two tasks $i$ and $j$. The operator is graphically shown in Fig.~\ref{Fig:Fig3-hong}.
	
\begin{figure}[htb]
		\begin{center}
			\subfigure[\small{}]{\psfig{file=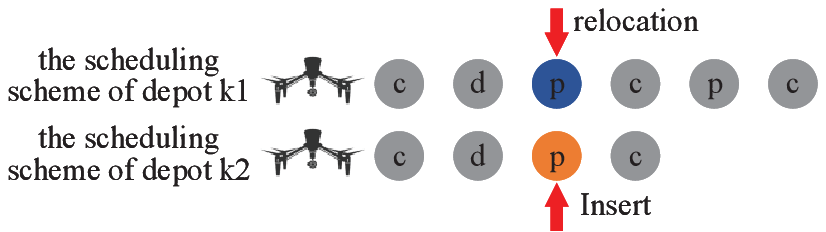,width=3.3in}}\\
			\subfigure[\small{}]{\psfig{file=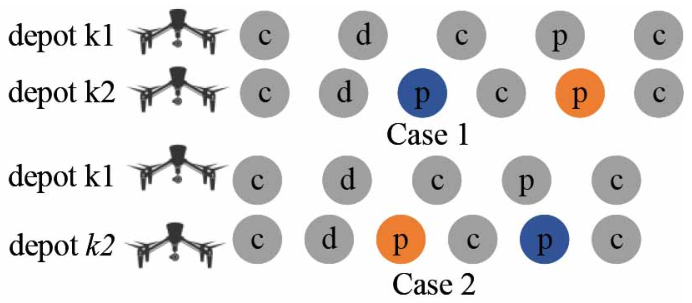,width=3.3in}}\\
			
		\end{center}
		\caption{Relocation. (a) Before Relocation. (b) After Relocation.}\label{Fig:Fig4-hong}
	\end{figure}

	b. 3-exchange: exchange three tasks assigned the same depot. The operator is similar to 2-exchange.

	c. 30\%-exchange: 2-exchange and 3-exchange are not suitable in dealing with problem of large-scale. We design 30\%-exchange operator to deal with scheduling with large-scale tasks. The operator is similar to 2-exchange: select 30\% of tasks assigned to a depot randomly, then exchange their position randomly.

	d. Relocation: select a task $i$ from all the tasks assigned to depot $k1$ randomly, and swap it to depot $k2$ for serving ($i \in PICKUP$). The operation of relocation is as follows: select a task $i$ from depot $k1$. Then, an insertion point is randomly selected in the existing allocation scheme of depot $k2$. Insert task $i$ into the position of the insertion point according to various constraints. The operator is graphically shown in Fig.~\ref{Fig:Fig4-hong}.

	\subsubsection{Improved variable neighborhood search algorithm (IVND) for task reallocation}

	Variable neighborhood search algorithm is an improved metaheuristic optimization algorithm, which can be used to solve combinatorial optimization problems. Here we design an improved variable neighborhood search algorithm (IVND) to refine allocation scheme: When the iteration times $L$ is not reached, we input an initial allocation scheme $G$ into the algorithm. Then one of the above six neighborhood transformation operators is chosen randomly to refine the allocation scheme $G$. A new solution $G'$ can be obtained for each cycle. Then, route planning scheme is generated based on the result of task reallocation scheme $S_0$. Also, we calculate the objective function value $f(S_0)$. According to Metropolis principle, we update the value of the current local optimal solution $S$. Repeat the above steps until the maximum number of iterations $L$ is reached. The pseudocode of the IVND is given as Algorithm 2.

	\begin{algorithm}
		\caption{\textit{IVND}}\label{PC:al2}
		\KwIn{allocation scheme $G$;\mbox{ } current objective function value $f(S)$;}
		\KwOut{$G'$;\mbox{ }$S$;}
		$i=1$\;
		
		\While{$i\le L$}
		{
			Generate new allocation scheme $G'$ from $G$ via one of the neighborhood transformation operators\;
			Generate new route planning $S_0$\;
			Calculate the cost of $S\to f(S_0)$\;
			$df=f(S_0)-f(S)$\;
			Generate a random number $\varepsilon\gets  rand(0,1)$
			
			\If{$df<0$}
			{
				$S\gets S_0,G\gets G',f(s)\gets f(S_0)$\;
			}
			
			\ElseIf{$exp(-df/t)\ge \varepsilon$}
			{
				$S\gets S_0,G\gets G',f(s)\gets f(S_0)$\;
			}
			$i=i+1$\;
		}
	\end{algorithm}

	\subsection{Route Planning}
	\subsubsection{Initial route planning}

	The initial task allocation scheme of each depot is disordered so that it cannot be directly executed by drones. Therefore, we design an elitist-based route planning algorithm (ERPA) to generate an initial route planning scheme based on the result of initial task allocation scheme. Previously, we have differentiated tasks into three types and divided them into three sets, including $DROP$, $PICKUP$, $PICK-DROP$.

	Since a drone can deliver one package each time, there are several rules for generating route planning of drones: firstly, tasks which belong to the $DROP$ set and the $PICK-DROP$ set will be completed by one drone to load package from depot and drop them to the right place; Secondly, tasks which belong to the $PICKUP$ set may be assigned to the drone which have been assigned task (the task must belong to the $DROP$ set) or a new drone to complete.

	According to the above rules, the route planning scheme of drones has the following rules: Firstly, the drone must start from a depot and finally return to the depot after completing all its tasks. Secondly, after the drone drops package to customer, it can choose to return to depot directly or go to the nearest customer to pick up package and then return to a depot. Finally, since drone can only carry one package each time, it must return to depot after picking up package. Thus, there are three kinds of possible route for drones (Fig.~\ref{Fig:Fig5-hong}):

	\begin{figure}[!htb]
		\begin{center}
			\subfigure{\psfig{file=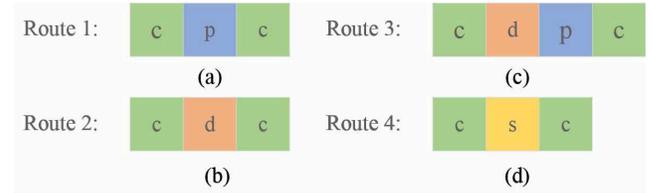, angle=0, width=3.3in}}
		\end{center}\vspace{-5mm}
		\caption{Possible route of drones (c: depot, d: task which need package delivery only, p: task which need package pickup only, s: task which need package delivery and pickup).}\label{Fig:Fig5-hong}
	\end{figure}
	
	(1) The drone starts from depot for completing the task of picking up package, and then returns to depot (see Fig.~\ref{Fig:Fig5-hong}(a));

	(2) The drone starts from a depot for completing the task of delivering package, and then returns to depot (see Fig.~\ref{Fig:Fig5-hong}(b));

	(3) The drone starts from depot for completing the task of dropping package, and then goes to the nearest customer to pick up package, finally returns to depot (see Fig.~\ref{Fig:Fig5-hong}(c));

	(4) If task $i$ need deliver and pick up package as well, the drone starts from depot for dropping package, and the reload package. After that, the drone returns to depot with package (see Fig.~\ref{Fig:Fig5-hong}(d));

\begin{figure}[!htb]
		\begin{center}
			\subfigure{\psfig{file=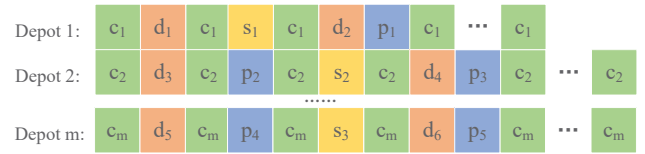, angle=0, width=3.3in}}
		\end{center}\vspace{-5mm}
		\caption{A route planning scheme of depots.}\label{Fig:Fig6-hong}
	\end{figure}

	In this paper, each depot and task have a unique index, so as to distinguish depots and different type of tasks. We index tasks firstly. Then, we index depots based on the index of tasks. For example, we assume that the number of tasks is $c$, and the number of depots is $m$, then the index of depots will be $c+1,c+2,\cdots,c+m$. As shown in Fig.~\ref{Fig:Fig6-hong}, the route planning of each depot can be freely combined by the above route of drones. Significantly, the route planning scheme of each depot are combined to form a complete scheduling scheme.

	\begin{algorithm}
		\caption{\textit{Elitist-based Route Planning Algorithm (ERPA)}}\label{PC:al3}
		\KwIn{task allocation schemes of depot $k$ $G_k$;\mbox{ }the number of depots $m$;\mbox{ }the number of tasks $c$}
		\KwOut{the initial route planning $S_0$;}
		
		\For{$k=1: m$}
		{
			Rearrange sequences of tasks in $G_k\to P_k$\;
			\For{$n=1:\mbox{ }$the number of tasks in $P_k$}
			{
				\If{task $n\in DROP\bigcup PICK-DROP$}
				{
					Arrange a drone to complete task $n$\;
				}
				
				\ElseIf{task $n\in PICK$}
				{
					\If{task $n-1\in DROP$}				
					{
						Generate a random number $\varepsilon\gets  rand(0,1)$\;
						\eIf{$\varepsilon\le 0.5$}
						{
							Arrange a drone to complete task $n$\;
						}
						{
							Assign task $n$ to the drone which have been arranged to complete task $n-1$\;
						}
					}
				}
			}
			Merge the schemes of each depot $\to S_0$\;
		}
	\end{algorithm}
	
	An elitist-based route planning algorithm (ERPA) are proposed to obtain an initial route planning scheme for depots: Firstly, for each depot $k$, we rearrange sequences of tasks based on the task allocation scheme $G_k$ and obtain an initial scheduled task set $P_k$. And for each task in the set $P_k$, if the task $n \in DROP \bigcup PICK-DROP $,an unassigned drone will be arranged to complete the task $n$. Otherwise, the task will be randomly assigned to an unassigned drone or a drone which has been arranged task $n-1$ (the task $n-1$ must belong to the $PICKUP$ set). Besides, drones must start from depots and return to depots after completing tasks. Then we will obtain multiple drones' route, as shown in Fig.~\ref{Fig:Fig5-hong}. All drones' routes of each depot are connected together and form the route planning of each depot. The route planning scheme can be regarded as scheduling sub-scheme. We obtain the overall scheduling scheme for all depots by merging all the sub-schemes. In order to ensure the quality of the initial solution, we construct multiple initial solutions by above method, and then select the best solution as $S_0$. The pseudocode of the ERPA is shown as Algorithm 3.

	\subsubsection{Adjustment of route planning scheme}\

	a. Repair solutions

\begin{figure}[!htb]
		\begin{center}
			\subfigure{\psfig{file=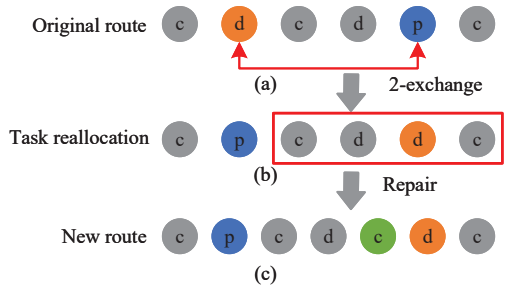, angle=0, width=3.3in}}
		\end{center}\vspace{-5mm}
		\caption{The operation of repairing.}\label{Fig:Fig7-hong}
	\end{figure}

	Task reallocation may make the original route planning scheme infeasible. For example, as shown in Fig.~\ref{Fig:Fig7-hong}, Fig.~\ref{Fig:Fig7-hong}(a) shows an initial route planning scheme for drones. However, after task reallocation by 2-exchange operator, the original route planning scheme is shown in Fig.~\ref{Fig:Fig7-hong}(b) and the scheme is obviously infeasible for drones to perform, because a drone cannot complete two tasks which belongs to the DROP set at once.Therefore, after task reallocation, the route planning scheme need to be repaired when the sub-route of the scheme does not conform to the four route types described in Fig.~\ref{Fig:Fig5-hong}. Also, route planning scheme will be repaired to be a combination of multiple routes of drones described in Fig.~\ref{Fig:Fig5-hong}. Fig.~\ref{Fig:Fig7-hong}(c) shows the repaired route planning scheme.

	b. Local search algorithm (LS)
	
\begin{figure}[!htb]
		\begin{center}
			\subfigure{\psfig{file=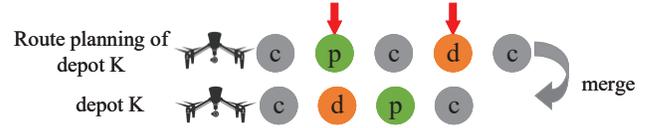, angle=0, width=3.3in}}
		\end{center}\vspace{-5mm}
		\caption{The operation of LS.}\label{Fig:Fig8-hong}
	\end{figure}

	In order to overcome the shortcoming of the above IVND algorithm and obtain the promising scheduling scheme, we design a local search algorithm (LS). LS refines the current scheduling scheme and obtains a better scheduling scheme. The operator is graphically shown in Fig.~\ref{Fig:Fig8-hong}: select two tasks which are assigned to two drones to serve originally and reassign them to one drone to serve.

	Specifically, the process of LS is as follows: we select a depot $k$ randomly, the current route planning scheme of depot $k$ is $s_0k$. Firstly, according to the task type of tasks, we add tasks in $s_{0k}$ into the set $s_{0k}-pick$, $s_{0k}-drop$, $s_{0k}-pd$. When the sets $s_{0k}-pick$ and $s_{0k}-drop$ are not empty, for tasks in the sets $s_{0k}-pick$, if $drone_{ik}$ is arranged to complete task $i$ only, then we add the task $i$ into the sets $temp1$. Repeat the similar operation for the set $s_{0k}-drop$. Finally, we select task $m$ and $n$ from the sets $temp1$ and $temp2$ respectively, and the above two tasks are reassigned to one drone for completing. The pseudocode of the LS is given as Algorithm 4.
	
	\begin{algorithm}
		\caption{\textit{local search algorithm(LS)}}\label{PC:al4}
		\KwIn{route planning $S_{0k}$ of the depot $k$;}
		\KwOut{$S_k$;}
		Classify tasks in $S_{0k}\to$ sets: $S_{0k}-pick$,$S_{0k}-drop$,$S_{0k}-pd$\;
		Initialize the unscheduled sets $temp1\gets \varnothing,temp2\gets \varnothing$\;
		
		\If{$S_{0k}-pick$ and $S_{0k}-drop$ $\neq 0$}
		{
			Calculate the number of tasks in $S_{0k}-pick$ $\to y1$\;
			\For{$i=1: y1$}
			{
				\If{$drone_{ik}$ complete task $i$ only}
				{Add $i$ to $temp1$\;}
			}
			Calculate the number of tasks in $S_{0k}-DROP$ $\to y2$\;
			\For{$i=1: y2$}
			{
				\If{$drone_{jk}$ complete task $j$ only}
				{Add $j$ to $temp2$\;}
			}
			Select $m\in temp1,n\in temp2$ randomly\;
			Reassign task $m,n$ to one drone to complete $\to S_k$\;
		}
	\end{algorithm}

	\section{Computational Experiments}
	
	In this section, we evaluate the effectiveness of SATO-IVND algorithm by comparing with six other heuristic algorithms and metaheuristic algorithms. In addition, we test the SATO-IVND algorithm in real scenes to prove its effectiveness. The proposed algorithm and other comparative algorithms are coded in python, and run on a PC computer with Core i5-8400 2.80GHz CPU, 8G memory, and Windows 10 operating system.
	
	\subsection{Experimental Setting}
	
	\definecolor{hl}{rgb}{0.75,0.75,0.75}
	\sethlcolor{hl}
	\begin{table}[htp]
		\caption{The parameter of sato-ivnd} \label{Table:table2} \vspace{-3mm}
		\footnotesize
		\renewcommand{\arraystretch}{1.5}
		\begin{center}
			\begin{tabular}{ll}
				\toprule
				parameters & value \\ \midrule
				Initial temperature $T_0$ & 1000 \\
				Terminating temperature $T_{end}$ & 1e-07 \\
				Cooling rate $q$ & 0.93 \\
				Maximum number of iterations $L$ & 20 \\
				Flying range (km) & 30 \\
				\bottomrule
			\end{tabular}
		\end{center}
	\end{table}

	Owing that a few studyies concern drop-pickup task scheduling, to the best of our knowledge, is currently no public benchmark for this problem . Therefore, we randomly generate thirteen instances whose number of tasks are 40, 60, 80, 100, 150, 200 respectively. Besides, numerous experiments are conducted in the simulation scenario and the realistic scenario to explore the performance of the proposed SATO-IVND.

	The simulation scenario is applied in a 50 km $\times$ 50 km area. Tasks are randomly distributed in the region. Suppose that all drones are homogenous. The weight of each package is a random number between 1kg and 8kg. Relevant parameter settings are shown in Table~\ref{Table:table2}. The parameters of the SATO-IVND are determined according to the relevant literature \cite{RN46} or via the trial-and-error way \cite{gong2018finding}.

	\subsection{Comparison with Other Algorithm}
	
	In order to verify the performance of the SATO-IVND algorithm, we compare SATO-IVND algorithm with other six heuristics and metaheuristics. We randomly generate thirteen instances where the number of tasks are 40, 60, 80, 100, 150, 200 respectively. Each algorithm runs 10 times to solve each instance. There are two reasons for the setting of this experiment: Firstly, it is proposed to verify the performance of the two-phase optimization approach in solving integrated scheduling problem of package drop-pickup considering m-drone, m-depot, m-customer; secondly, it is aimed to examine the effectiveness of the initial solution generation mechanism and neighborhood structures of SATO-IVND.

	In order to verify the performance of the two-phase optimization approach in solving the original problems, ERPA and three efficient algorithms, namely ALNS \cite{ropke2006adaptive,masson2013adaptive}, improved genetic algorithm (IGA) \cite{2008Research,bae2007integrated}and LNS \cite{gendreau2008tabu}, are designed for integrated scheduling problem and chosen as comparison algorithms. $k$-means \cite{2008Research} is adopted in IGA and LNS to generate the initial task allocation scheme, while ALNS randomly generate the initial task allocation scheme. Meanwhile, in order to ensure that the good individuals of the parents are not lost during the evolutionary process, the elite strategy is adopted in IGA to retain the optimal individuals of the parents. Both ALNS and LNS algorithms apply damage and repair operators to iteratively optimize until the termination conditions are met. For LNS, the neighborhood operators such as 2-opt, simple relocate, and swap \cite{gendreau2008tabu} are used for route planning, while ALNS integrates adaptive mechanism and worst-remove operator on the basis of multiple neighborhood operators  \cite{gendreau2008tabu}.

	For testing the effectiveness of initial solution generation mechanism and neighborhood structure of SATO-IVND, we compare SATO-IVND with two other algorithms including SA-IVND and SATO-IVND without LS algorithm (SATO-VND1). SA-IVND indicates that the initial task allocation scheme is generated randomly. Besides, the proposed IVND and LS algorithms are used to find the refine solution.

		
	\definecolor{hl}{rgb}{0.75,0.75,0.75}
	\sethlcolor{hl}
	\begin{table*}[htp]
		\caption{The results of each instance generated by sato-ivnd and other six lgorithms} \label{Table:table3} 
		\footnotesize
		\renewcommand{\arraystretch}{1.5}
		\begin{center}
			\begin{tabular}{@{}ccccccccccccccccc@{}}
					\toprule
					\multirow{2}{*}{Instance} & \multirow{2}{*}{Num-C} & \multirow{2}{*}{Num-D} & \multicolumn{2}{c}{SATO-IVND} & \multicolumn{2}{c}{ERPA} & \multicolumn{2}{c}{SATO-VND1} & \multicolumn{2}{c}{SA-IVND} & \multicolumn{2}{c}{LNS} & \multicolumn{2}{c}{IGA} & \multicolumn{2}{c}{ALNS} \\ \cline{4-17}
					&  &  & Cost & Time & Cost & Time & Cost & Time & Cost & Time & Cost & Time & Cost & Time & Cost & Time \\ \midrule
					C1 & 40 & 5 & 50 & 6.9 & 57 & 0.05 & 53.9 & 6.51 & 58.7 & 6.75 & 54.6 & 36.15 & 54.6 & 36.15 & 51.2 & 18 \\
					C2 & 60 & 5 & 80.1 & 10.33 & 101.3 & 0.08 & 93.7 & 10.02 & 103 & 10.44 & 95.8 & 53.92 & 89.6 & 53.92 & 85.9 & 27.18 \\
					C3 & 80 & 5 & 108.1 & 13.19 & 134.5 & 0.1 & 123.3 & 12.55 & 136.6 & 13.18 & 123 & 70.07 & 116.9 & 70.07 & 121.4 & 33.71 \\
					C4 & 100 & 5 & 132.4 & 17.48 & 164.6 & 0.14 & 187.2 & 17.02 & 171.8 & 17.66 & 165.7 & 91.66 & 153.2 & 91.66 & 191.6 & 44.93 \\
					C5 & 150 & 5 & 209.7 & 26.68 & 268.4 & 0.21 & 248.1 & 25.05 & 283.5 & 26.26 & 261.3 & 137.76 & 252.7 & 137.76 & 250.2 & 69.04 \\
					C6 & 200 & 5 & 284.5 & 36.89 & 363.5 & 0.29 & 339.6 & 35.71 & 371.3 & 36.63 & 357.9 & 185.47 & 350.1 & 185.47 & 334.3 & 95.86 \\
					C7 & 40 & 2 & 72.5 & 6.6 & 90.5 & 0.05 & 84.8 & 6.27 & 72.6 & 6.62 & 84.1 & 36.06 & 83.9 & 36.06 & 82.1 & 17.39 \\
					C8 & 40 & 4 & 64.1 & 6.62 & 72.6 & 0.05 & 67.9 & 6.5 & 74 & 6.7 & 70.6 & 35.41 & 67.3 & 35.41 & 66.6 & 18.14 \\
					C9 & 60 & 3 & 102.7 & 10.49 & 137.2 & 0.08 & 121.7 & 10.05 & 113.2 & 10.11 & 122.9 & 53.45 & 126.3 & 53.45 & 112.6 & 26.83 \\
					C10 & 80 & 4 & 126.9 & 13.27 & 155.5 & 0.1 & 142.7 & 12.58 & 142 & 13.32 & 150.6 & 71.11 & 142.5 & 71.11 & 142.1 & 34.73 \\
					C11 & 100 & 10 & 90.2 & 17.51 & 108.6 & 0.14 & 97.5 & 16.88 & 168.2 & 17.56 & 106.7 & 90.5 & 95.8 & 90.5 & 101.2 & 45.82 \\
					C12 & 150 & 7 & 164.5 & 26.47 & 209.4 & 0.21 & 196.7 & 25.93 & 270.6 & 26.57 & 199.3 & 137.35 & 191.8 & 137.35 & 189.6 & 68.72 \\
					C13 & 200 & 10 & 193.9 & 36.9 & 247.3 & 0.3 & 228.5 & 35.72 & 409.2 & 36.88 & 243.3 & 183.38 & 231 & 183.38 & 225.6 & 96.02 \\ \bottomrule
			\end{tabular}
		\end{center}
	\end{table*}
	
	SATO-IVND and the above six algorithms run 10 times to solve the thirteen instances. The number of tasks in different instances is donated by \textit{Num-C}. The number of depots in different instances is donated by \textit{Num-D}. The results are compared in terms of cost. The maximum value, the minimum value and the average value of 10 experimental results are presented. All the results are shown in Table~\ref{Table:table3}.

\begin{figure}[!htb]
	\begin{center}
		\subfigure{\psfig{file=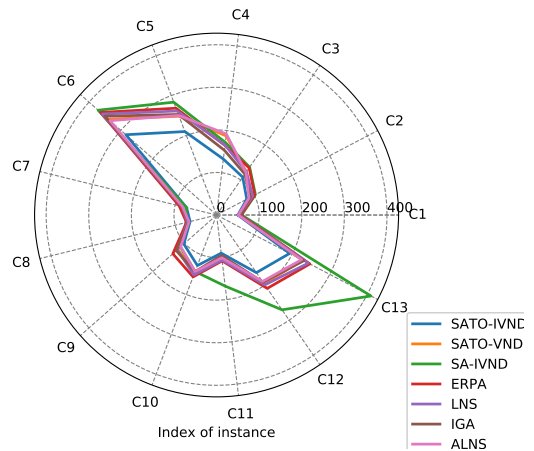, angle=0, width=3.3in}}
	\end{center}\vspace{-5mm}
	\caption{Results of experimental instances generated by different algorithms.}\label{Fig:Fig9-hong}
	\end{figure}

	 It can observe in Table~\ref{Table:table3} that the proposed SATO-IVND is superior to other six algorithms in terms of the solution quality. To clearly compare the performance of SATO-IVND and other six algorithms on the different instances, the values of the objective function are presented in Fig.~\ref{Fig:Fig9-hong}. As shown in Fig.~\ref{Fig:Fig9-hong}, with the increasing number of tasks, the results of SATO-IVND are always better than other six algorithms. Especially, the cost of instances \textit{C}6 and \textit{C}13 increases suddenly, which is due to the significant increase of the number of tasks.

	In order to further compare the result obtained by the other six algorithms and SATO-IVND for the same instance, we use Gap to describe the differences between SATO-IVND and other algorithms, which can be calculated by the following formula:
	
	\begin{equation}\label{eq18}
	\begin{array}{cc}
	Gap=\frac{S_i-S_{SATO_IVND}}{S_i}
	\end{array}
	\end{equation}
	where $S_i$ means the satisfactory cost found by algorithm $i$, and $i$ represents the one of ALNS, IGA, LNS, SA-IVND, ERPA and SATO-VND1. $S_{SATO-IVND}$ denotes the satisfactory solution found by SATO-IVND. The gap of the objective value generated by six algorithms compared to SATO-IVND is illustrated in Table~\ref{Table:table4}.

	From Table~\ref{Table:table4}, it shows that SATO-IVND outperforms other six heuristics and metaheuristics in solving thirteen instances, with the average gap value are from 11.95\% to 23.24\%. It is also worth noting that the Gap values are increasing with the increasing number of tasks. The results obtained by SATO-IVND are very close to the results obtained by other six algorithms when the size of tasks is small, but the effectiveness of other six algorithms decrease quickly when dealing with large-scale tasks. Among them, the Gap between the results obtained by ALNS and the results obtained by SATO-IVND is even more than 30\%. Overall, SATO-IVND shows visible performance in solving large-scale instances.

	Compared with LNS, as shown in Table~\ref{Table:table4}, SATO-IVND reduces the values of the objective function by 15.82\% on average. The 8.42\% minimum Gap and the 20.52\% maximum Gap indicate that SATO-IVND improves the solution quality enormously than LNS. Moreover, the minimum gap value of IGA is 4.80\% and the maximum gap value is 18.74\%. The minimum gap value of ALNS is 2.33\% and the maximum gap value is 30.90\%. The same conclusion can be obtained from the results of Gap value of IGA and ALNS that the proposed SATO-IVND is superior to IGA and ALNS.

	\definecolor{hl}{rgb}{0.75,0.75,0.75}
	\sethlcolor{hl}
	\begin{table*}[htp]
		\caption{Gap and c.v. of algorithms} \label{Table:table4} \vspace{-3mm}
		\footnotesize
		\renewcommand{\arraystretch}{1.5}
		\begin{center}
			\begin{tabular}{cccccccccccc}
				\toprule
				\multirow{2}{*}{Instance} & \multirow{2}{*}{Num-C} & \multirow{2}{*}{Num-D} & \multicolumn{6}{c}{Gap   (\%)} &  & \multicolumn{2}{c}{C.V(\%)} \\ \cline{4-9} \cline{11-12}
				&  &  & ERPA & SATO-VND1 & SA-IVND & LNS & IGA & ALNS & \multicolumn{1}{c}{} & SA-IVND & SATO-IVND \\ \midrule
				C1 & 40 & 5 & 12.27\% & 7.33\% & 14.90\% & 8.42\% & 8.45\% & 2.33\% &  & 13.8 & 0.7 \\
				C2 & 60 & 5 & 20.93\% & 14.52\% & 22.25\% & 16.38\% & 10.60\% & 6.75\% &  & 17.4 & 1 \\
				C3 & 80 & 5 & 19.66\% & 12.31\% & 20.87\% & 12.11\% & 7.52\% & 11.00\% &  & 16.9 & 1 \\
				C4 & 100 & 5 & 19.56\% & 29.28\% & 22.94\% & 20.06\% & 13.55\% & 30.90\% &  & 8.5 & 1 \\
				C5 & 150 & 5 & 21.86\% & 15.47\% & 26.02\% & 19.75\% & 17.02\% & 16.17\% &  & 4 & 1.4 \\
				C6 & 200 & 5 & 21.73\% & 16.24\% & 23.38\% & 20.52\% & 18.74\% & 14.89\% &  & 10.9 & 1.3 \\
				C7 & 40 & 2 & 19.87\% & 14.45\% & 0.16\% & 13.75\% & 13.56\% & 11.69\% &  & 0.6 & 0.5 \\
				C8 & 40 & 4 & 11.72\% & 5.66\% & 13.46\% & 9.21\% & 4.80\% & 3.89\% &  & 5.5 & 0.9 \\
				C9 & 60 & 3 & 25.17\% & 15.66\% & 9.32\% & 16.47\% & 18.70\% & 8.81\% &  & 7.1 & 3 \\
				C10 & 80 & 4 & 18.38\% & 11.09\% & 10.64\% & 15.76\% & 10.95\% & 10.73\% &  & 6 & 1.2 \\
				C11 & 100 & 10 & 16.96\% & 7.55\% & 46.40\% & 15.47\% & 5.83\% & 10.91\% &  & 35 & 0.5 \\
				C12 & 150 & 7 & 21.45\% & 16.35\% & 39.20\% & 17.45\% & 14.21\% & 13.22\% &  & 15.4 & 1.5 \\
				C13 & 200 & 10 & 21.61\% & 15.16\% & 52.62\% & 20.33\% & 16.06\% & 14.05\% & \multicolumn{1}{c}{} & 21.0 & 0.9 \\ \midrule \multicolumn{3}{c}{Min} & 11.72\% & 5.66\% & 0.16\% & 8.42\% & 4.80\% & 2.33\% &  & 0.6 & 0.5 \\
				\multicolumn{3}{c}{Max} & 25.17\% & 29.28\% & 52.62\% & 20.52\% & 18.74\% & 30.90\% &  & 35 & 3 \\
				\multicolumn{3}{c}{Average} & 19.32\% & 13.93\% & 23.24\% & 15.82\% & 12.31\% & 11.95\% & \multicolumn{1}{c}{} & 12.5 & 1.1 \\ \bottomrule
			\end{tabular}
		\end{center}
	\end{table*}

	Relative to ERPA, SATO-IVND reduces the values of the objective function by 19.32\% on average. The 11.72\% minimum Gap and the 25.17\% maximum Gap indicate that the interactive iteration in SATO-IVND can effectively refine the initial scheme and find the high-quality scheduling scheme.

	Compared with SATO-VND1, SATO-IVND reduces the values of the objective function by 13.93\% on average. The reason for the Gap between SATO-IVND and SATO-VND1 is that SATO-VND1 does not adopt the proposed LS algorithm incorporated in SATO-VND, resulting in SATO-VND1 premature convergence on local optimum.

	The average Gap of SA-IVND is 23.24\%, and the 0.16\% minimum Gap and the 52.62\% maximum Gap indicate that SATO-IVND is obviously superior to SA-IVND in terms of cost. The different initial allocation scheme generation mechanism leads to the Gap between SATO-IVND and SA-IVND. That is, the quality of the initial allocation scheme will affect the solution quality of algorithms.
	
	\begin{figure}[!htb]
		\begin{center}
			\subfigure{\psfig{file=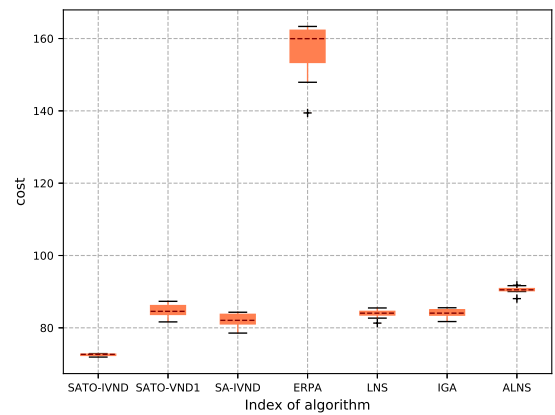, angle=0, width=3.3in}}
		\end{center}\vspace{-5mm}
		\caption{Results of experimental instances C8 generated by different algorithms.}\label{Fig:Fig10-hong}
	\end{figure}
	
	The robustness of the proposed SATO-IVND and SA-IVND is quantified by the coefficient of variation of cost, which is the ratio of standard deviation to mean of cost for each instance. The results are covered in Table~\ref{Table:table4}. The coefficient of variation of cost is donated by C.V in Table~\ref{Table:table4}. As shown in Table~\ref{Table:table4}, the robustness of SATO-IVND is significantly better than that of SA-IVND. Besides, we present the calculation results of instance \textit{C}8 at Fig.~\ref{Fig:Fig10-hong}. It can be seen from Fig.~\ref{Fig:Fig10-hong} that SATO-IVND is obviously superior to the other six algorithms in terms of robustness and optimization effect.
	
	\begin{figure}[!htb]
		\begin{center}
			\subfigure{\psfig{file=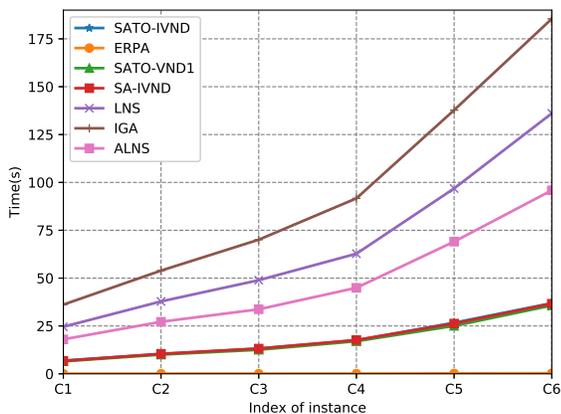, angle=0, width=3.3in}}
		\end{center}\vspace{-5mm}
		\caption{The running time of algorithms for instance C1-C6.}\label{Fig:Fig11-hong}
	\end{figure}
	
	In terms of time consumption, as shown in Table~\ref{Table:table4} and Fig.~\ref{Fig:Fig11-hong}, ERPA is better than the other six algorithms, followed by SA-IVND, SATO-IVND, SATO-VND1. In contrast, with the increasing number of tasks, the running time of IGA, LNS, ALNS increase rapidly. Among them, IGA is the most time-consuming, followed by LNS and ALNS. Specially, with the increasing number of tasks, IGA significantly spends more time than other algorithms to find the promising solution. The difference between LNS and ALNS is that LNS generates the initial allocation scheme based on the $k$-means algorithm, while ALNS generates the initial allocation scheme randomly. From the comparison of the above two algorithms, the $k$-means algorithm can greatly improve the quality of solution in less time consumption.

	However, LNS still has the shortcoming of premature convergence compared with SATO-IVND. SATO-IVND is much better than LNS in cost optimization because it adopts a variety of neighborhood operators proposed in this paper. Meanwhile, the running time of SATO-IVND is significantly shorter than that of LNS, indicating that SATO-IVND can find a satisfactory solution in a shorter time than LNS. The running time of algorithms for each instance are shown in Table~\ref{Table:table4} and Fig.~\ref{Fig:Fig11-hong}.

	ERPA is obviously better than SATO-IVND in time consumption, but the solutions of SATO-IVND are obviously better than ERPA. It proves that SATO-IVND can improve the quality of solution to a great extent, and ERPA is easy to converge to the local optima. By contrast, ERPA may be applied to situations that are very time-sensitive but do not require high solution quality.
	Despite that SATO-IVND and SATO-VND1 are based on two-phase optimization approach, SATO-IVND surpasses SATO-VND1 with regard to profit ratio due to the proposed LS algorithm. Significantly, with regard to calculation time, calculation time of SATO-IVND is very close to SATO-VND1, indicating that the proposed LS algorithm can improve the quality of solution with very little time consumption. Additionally, the calculation time of SA-IVND is close to SATO-IVND, as shown in Table~\ref{Table:table4}. However, SATO-IVND is superior to SA-IVND in terms of cost, which indicates that the initial solution generation mechanism can effectively improve the quality of solution in a reasonable time.

	In short, the two-phase optimization approach has a good application prospect in solving integrated scheduling problem considering \textit{m}-drone, \textit{m}-depot, \textit{m}-customer constrained by drop-pickup. The two-phase optimization approach for SATO-IVND can produce high-quality scheduling scheme in a reasonable time. Considering the balance between computation and solution quality, SATO-IVND is very suitable for large-scale task scheduling problems considering \textit{m}-drone, \textit{m}-depot, \textit{m}-customer. In contrast, ERPA can be applied in time-sensitive scenarios at the expense of solution quality.

	\subsection{Experiment in a Realistic Scenario}
	
\begin{figure}[!htb]
		\begin{center}
			\subfigure{\psfig{file=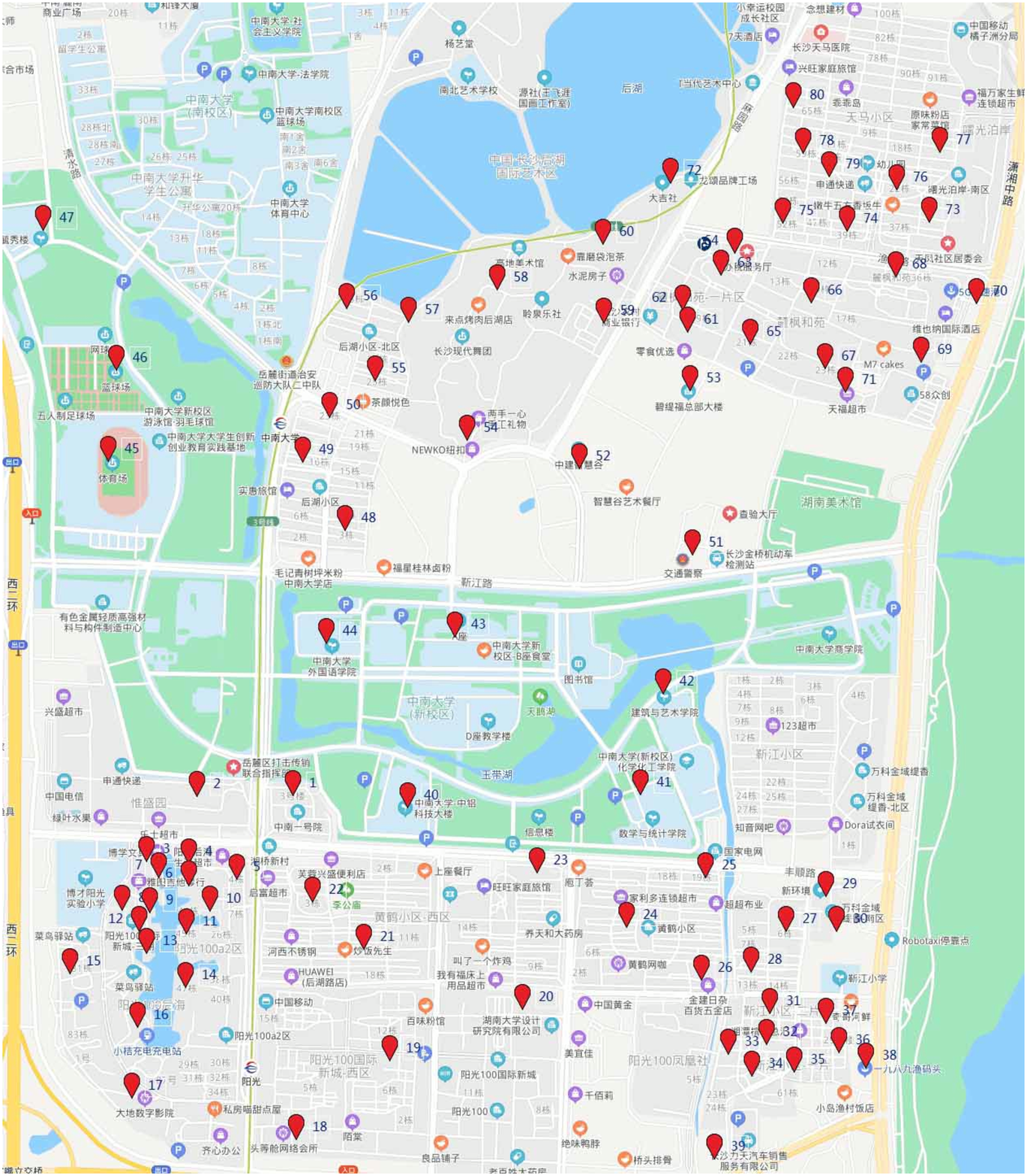, angle=0, width=3.3in}}
		\end{center}\vspace{-5mm}
		\caption{Illustration of 80 tasks in Changsha City (red dot: the location of tasks).}\label{Fig:Fig12-hong}
	\end{figure}

	In order to further verify the effectiveness of SATO-IVND in solving integrated scheduling problem considering \textit{m}-drone, \textit{m}-depot, \textit{m}-customer constrained by drop-pickup, the following experiment is conducted under a realistic scenario. In this paper, the related region is in Yuelu District, Changsha City, Hunan Province. We select 80 tasks completed by drones, as illustrated in Fig.~\ref{Fig:Fig12-hong}. In addition, we set 5 depots in this area. The weight of each package is a random number within 1kg-8kg.
	
	\definecolor{hl}{rgb}{0.75,0.75,0.75}
	\sethlcolor{hl}
	\renewcommand\arraystretch{1.5}
	\begin{table}[htp]
		\caption{Scheduling scheme generated by sato-ivnd} \label{Table:table6} 
		\footnotesize
		\renewcommand{\arraystretch}{1.5}
		\begin{center}
			\begin{tabular}{cc}
				\toprule
				\makecell{Depot \\ NO.} & Scheduling Scheme \\
				\midrule
				01        & \makecell{01$\rightarrow$44$\rightarrow$48$\rightarrow$01$\rightarrow$50$\rightarrow$01$\rightarrow$56$\rightarrow$55$\rightarrow$01$\rightarrow$57$\rightarrow$54$\rightarrow$01$\rightarrow$\\49$\rightarrow$43$\rightarrow$01$\rightarrow$47$\rightarrow$45$\rightarrow$01$\rightarrow$46$\rightarrow$01} \\  \hline
				02        & \makecell{02$\rightarrow$36$\rightarrow$31$\rightarrow$02$\rightarrow$24$\rightarrow$02$\rightarrow$27$\rightarrow$02$\rightarrow$20$\rightarrow$23$\rightarrow$02$\rightarrow$32$\rightarrow$\\28$\rightarrow$02$\rightarrow$33$\rightarrow$34$\rightarrow$02$\rightarrow$38$\rightarrow$35$\rightarrow$02$\rightarrow$37$\rightarrow$30$\rightarrow$02$\rightarrow$29$\rightarrow$\\02$\rightarrow$39$\rightarrow$02$\rightarrow$42$\rightarrow$25$\rightarrow$02$\rightarrow$26$\rightarrow$02$\rightarrow$41$\rightarrow$02}    \\ \hline
				03        & \makecell{03$\rightarrow$52$\rightarrow$51$\rightarrow$03$\rightarrow$59$\rightarrow$03$\rightarrow$64$\rightarrow$65$\rightarrow$03$\rightarrow$72$\rightarrow$63$\rightarrow$03$\rightarrow$\\60$\rightarrow$03$\rightarrow$62$\rightarrow$03$\rightarrow$53$\rightarrow$03$\rightarrow$58$\rightarrow$03$\rightarrow$61$\rightarrow$03}    \\ \hline
				04        & \makecell{04$\rightarrow$12$\rightarrow$8$\rightarrow$04$\rightarrow$22$\rightarrow$04$\rightarrow$17$\rightarrow$15$\rightarrow$04$\rightarrow$9$\rightarrow$11$\rightarrow$04$\rightarrow$13\\$\rightarrow$04$\rightarrow$2$\rightarrow$10$\rightarrow$04$\rightarrow$16$\rightarrow$14$\rightarrow$04$\rightarrow$7$\rightarrow$3$\rightarrow$04$\rightarrow$18$\rightarrow$04$\rightarrow$19\\$\rightarrow$04$\rightarrow$4$\rightarrow$04$\rightarrow$40$\rightarrow$5$\rightarrow$04$\rightarrow$1$\rightarrow$04$\rightarrow$6$\rightarrow$04$\rightarrow$21$\rightarrow$04} \\ \hline
				05        & \makecell{05$\rightarrow$67$\rightarrow$71$\rightarrow$05$\rightarrow$69$\rightarrow$70$\rightarrow$05$\rightarrow$73$\rightarrow$05$\rightarrow$76$\rightarrow$05$\rightarrow$78$\rightarrow$\\75$\rightarrow$05$\rightarrow$80$\rightarrow$05$\rightarrow$79$\rightarrow$74$\rightarrow$05$\rightarrow$77$\rightarrow$68$\rightarrow$05$\rightarrow$66$\rightarrow$05} \\  \bottomrule
			\end{tabular}
		\end{center}
	\end{table}
	
	\begin{figure}[!htb]
		\begin{center}
			\subfigure{\psfig{file=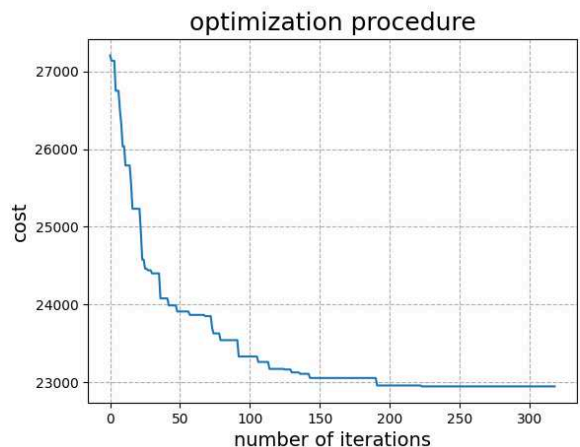, angle=0, width=3.3in}}
		\end{center}\vspace{-5mm}
		\caption{Convergence curve of SATO-IVND.}\label{Fig:Fig13-hong}
	\end{figure}
	
	The scheduling scheme generated by SATO-IVND is shown in Table~\ref{Table:table6}, and the cost convergence curve for SATO-IVND is shown in Fig.~\ref{Fig:Fig13-hong}. The index of depot is donated by Depot NO. The running time is 12.91s, and the cost of the satisfactory solution is reduced by 15.8\% compared with the initial solution.

	As can be seen from Fig.~\ref{Fig:Fig13-hong}, the cost of the scheduling scheme generated by SATO-IVND decreases rapidly within 50 iterations, indicating that SATO-IVND has strong optimization ability in a short time. The Metropolis principle and IVND operators and LS operators avoid premature convergence of SATO-IVND. The algorithm converges to the promising solution in 225 iterations.

	\section{Conclusions and Future Work}
	
	In this paper, we propose a novel drone package pickup and delivery mode and system to deal with problems existing in last-mile delivery. Meanwhile, a two-phase optimization approach (SATO-IVND) is proposed to solve the integrated scheduling problem considering \textit{m}-drone, \textit{m}-depot, \textit{m}-customer constrained by drop-pickup, and an effective scheduling scheme for drones is obtained. Different from the exact algorithm, SATO-IVND can find a high-quality scheduling scheme in a reasonable time, which can be used to solve large-scale problems and applied to real scenes. We decompose the integrated scheduling problem into two stages: task allocation and route planning. In the first phase, we use the $k$-means algorithm to generate the initial allocation scheme, and we design the IVND algorithm which includes six operators to reallocate tasks. In the second phase, the ERPA algorithm is proposed to generate the initial scheduling scheme. Meanwhile, we design the LS algorithm to refine the current scheduling scheme. Finally, the high-quality scheduling scheme is obtained through repeated iteration.
	
	The effectiveness of SATO-IVND is verified by numerous experiments, from which we can clearly draw the following conclusions: Firstly, compared with six heuristic algorithms and metaheuristic algorithms, it can be seen that the SATO-IVND is superior to the other six heuristic and metaheuristic algorithms in solving quality and time efficiency. Secondly, the experiment in the real scene shows that SATO-IVND can generate high-quality scheduling schemes in a reasonable time.
	
	The drone pick-up and delivery system proposed in this paper is a novel mode to solve the last-mile distribution problem. The drone pick-up and delivery mode and system is worth further researching in the future. For example, considering the time window, relevant advanced algorithms will be designed to produce a satisfactory scheduling scheme in a short time, contributing to the application of drone in the real-time pick-up and delivery platform \cite{wang2019vehicle}. In addition, this research can also be extended to the setting of charging station , the heterogeneous problem of drone \cite{murray2020multiple}, etc.

	%
	%

	\ifCLASSOPTIONcaptionsoff
	\newpage
	\fi

	\bibliographystyle{IEEEtran}
	\bibliography{IEEEabrv,mybib1}

\begin{thebibliography}{10}
\providecommand{\url}[1]{#1}
\csname url@samestyle\endcsname
\providecommand{\newblock}{\relax}
\providecommand{\bibinfo}[2]{#2}
\providecommand{\BIBentrySTDinterwordspacing}{\spaceskip=0pt\relax}
\providecommand{\BIBentryALTinterwordstretchfactor}{4}
\providecommand{\BIBentryALTinterwordspacing}{\spaceskip=\fontdimen2\font plus
\BIBentryALTinterwordstretchfactor\fontdimen3\font minus
  \fontdimen4\font\relax}
\providecommand{\BIBforeignlanguage}[2]{{%
\expandafter\ifx\csname l@#1\endcsname\relax
\typeout{** WARNING: IEEEtran.bst: No hyphenation pattern has been}%
\typeout{** loaded for the language `#1'. Using the pattern for}%
\typeout{** the default language instead.}%
\else
\language=\csname l@#1\endcsname
\fi
#2}}
\providecommand{\BIBdecl}{\relax}
\BIBdecl

\bibitem{2014Hierarchical}
T.~Aized and J.~S. Srai, ``Hierarchical modelling of last mile logistic
  distribution system,'' \emph{International Journal of Advanced Manufacturing
  Technology}, vol.~70, no. 5-8, pp. 1053--1061, 2014.

\bibitem{RN02}
A.~Meola, ``Shop online and get your items delivery by a drone delivery
  service: the future {Amazon} and domino’s have envisioned for {US},''
  \url{https://www.businessinsider.com/delivery-drones-market-service-2017-7},
  7 2017.

\bibitem{RN03}
A.~Hern, ``{DHL} launches first commercial drone ’parcelcopter’ delivery
  service,''
  \url{https://www.theguardian.com/technology/2014/sep/25/german-dhl-launches-first
  commercial-drone-delivery-service}, 9 2014.

\bibitem{RN04}
N.~Shields, ``China’s largest courier is starting drone deliveries,''
  https://www.businessinsider.com/chinas-
  largest-courier-to-start-drone-deliveries-2018-4.

\bibitem{RN06}
K.~research institute, ``Unmanned delivery ﬁeld research report,''
  http://www.199it.com/archives/1072522.html.

\bibitem{2020Joint}
M.~Salama and S.~Srinivas, ``Joint optimization of customer location clustering
  and drone-based routing for last-mile deliveries,'' \emph{Transportation
  Research Part C Emerging Technologies}, vol. 114, pp. 620--642, 2020.

\bibitem{2021A}
Y.~Choi and P.~M. Schonfeld, ``A comparison of optimized deliveries by drone
  and truck,'' \emph{Transportation Planning and Technology}, vol.~44, no.~3,
  pp. 319--336, 2021.

\bibitem{2017Vehicle}
H.~J. M. G.~G. Dorling, K. and S.~Magierowski, ``Vehicle routing problems for
  drone delivery,'' \emph{IEEE Transactions on Systems, Man, and Cybernetics:
  Systems}, vol.~47, no.~1, pp. 70--85, 2017.

\bibitem{2018Persistent}
B.~D. Song, K.~Park, and J.~Kim, ``Persistent {UAV} delivery logistics: {MILP}
  formulation and efficient heuristic,'' \emph{Computers \& Industrial
  Engineering}, vol. 120, pp. 418--428, 2018.

\bibitem{murray2015flying}
C.~C. Murray and A.~G. Chu, ``The flying sidekick traveling salesman problem:
  Optimization of drone-assisted parcel delivery,'' \emph{Transportation
  Research Part C: Emerging Technologies}, vol.~54, pp. 86--109, 2015.

\bibitem{tu2018traveling}
D.~N.~T. Tu, Phan~Anh and P.~Q. Dung, ``Traveling salesman problem with
  multiple drones,'' in \emph{Proceedings of the Ninth International Symposium
  on Information and Communication Technology}, 2018, pp. 46--53.

\bibitem{shima2006multiple}
T.~Shima, S.~J. Rasmussen, A.~G. Sparks, and K.~M. Passino, ``Multiple task
  assignments for cooperating uninhabited aerial vehicles using genetic
  algorithms,'' \emph{Computers \& Operations Research}, vol.~33, no.~11, pp.
  3252--3269, 2006.

\bibitem{2018Integrated}
A.~M. Ham, ``Integrated scheduling of m-truck, m-drone, and m-depot constrained
  by time-window, drop-pickup, and m-visit using constraint programming,''
  \emph{Transportation Research Part C Emerging Technologies}, vol.~91, pp.
  1--14, 2018.

\bibitem{2018A}
C.~W. T. G. H. Z. H. Y. Y. L. W.~Y. Jia, Yahui and Z.~Jun, ``A dynamic logistic
  dispatching system with set-based particle swarm optimization,'' \emph{IEEE
  Transactions on Systems, Man, and Cybernetics: Systems}, vol.~48, no.~9, pp.
  1607--1621, 2018.

\bibitem{2019An}
C.~T. M. L.~Z. Wang, X. and S.~Shao, ``An effective local search algorithm for
  the multi-depot cumulative capacitated vehicle routing problem,'' \emph{IEEE
  Transactions on Systems, Man, and Cybernetics: Systems}, vol.~50, no.~12, pp.
  4948--4958, 2019.

\bibitem{ancele2021toward}
Y.~Ancele, M.~H. H{\`a}, C.~Lersteau, D.~B. Matellini, and T.~T. Nguyen,
  ``Toward a more flexible {VRP} with pickup and delivery allowing
  consolidations,'' \emph{Transportation Research Part C: Emerging
  Technologies}, vol. 128, pp. 1--21, 2021.

\bibitem{das2020synchronized}
D.~N. Das, R.~Sewani, J.~Wang, and M.~K. Tiwari, ``Synchronized truck and drone
  routing in package delivery logistics,'' \emph{IEEE Transactions on
  Intelligent Transportation Systems}, vol.~22, no.~9, 2021.

\bibitem{2019Review}
X.~Ren, H.~Hui, Y.~Shaowei, S.~Feng, and L.~Gongqian, ``Review on vehicle-{UAV}
  combined delivery problem,'' \emph{Kongzhi Yu Juece}, pp. 1--15, 2020.

\bibitem{mathew2015planning}
N.~Mathew, S.~L. Smith, and S.~L. Waslander, ``Planning paths for package
  delivery in heterogeneous multirobot teams,'' \emph{IEEE Transactions on
  Automation Science and Engineering}, vol.~12, no.~4, pp. 1298--1308, 2015.

\bibitem{karak2019hybrid}
A.~Karak and K.~Abdelghany, ``The hybrid vehicle-drone routing problem for
  pick-up and delivery services,'' \emph{Transportation Research Part C:
  Emerging Technologies}, vol. 102, pp. 427--449, 2019.

\bibitem{wang2019expressway}
C.~Wang and H.~Lan, ``An expressway based {TSP} model for vehicle delivery
  service coordinated with {Truck+UAV},'' in \emph{2019 IEEE International
  Conference on Systems, Man and Cybernetics (SMC)}.\hskip 1em plus 0.5em minus
  0.4em\relax IEEE, 2019, pp. 307--311.

\bibitem{2017Same}
I.~Dayarian, M.~Savelsbergh, and J.~P. Clarke, ``Same-day delivery with drone
  resupply,'' vol.~54, no.~1, pp. 229--249, 2020.

\bibitem{chen2021clustering}
J.~Chen, C.~Du, Y.~Zhang, P.~Han, and W.~Wei, ``A clustering-based coverage
  path planning method for autonomous heterogeneous uavs,'' \emph{IEEE
  Transactions on Intelligent Transportation Systems}, 2021.

\bibitem{deng2020two}
M.~Deng, B.~Liu, S.~Li, R.~Du, G.~Wu, H.~Li, and L.~Wang, ``A two-phase
  coordinated planning approach for heterogeneous earth-observation resources
  to monitor area targets,'' \emph{IEEE Transactions on Systems, Man, and
  Cybernetics: Systems}, pp. 1--16, 2020.

\bibitem{2020Iterative}
L.~Huan, L.~Xiamiao, W.~Guohua, F.~Mingfeng, W.~Rui, G.~Liang, and W.~Pedrycz,
  ``An iterative two-phase optimization method based on divide and conquer
  framework for integrated scheduling of multiple {UAVs},'' \emph{IEEE
  Transactions on Intelligent Transportation Systems}, vol.~22, no.~9, pp. 5926
  -- 5938, 2021.

\bibitem{ma2021unsupervised}
S.~Ma, W.~Guo, R.~Song, and Y.~Liu, ``Unsupervised learning based coordinated
  multi-task allocation for unmanned surface vehicles,'' \emph{Neurocomputing},
  vol. 420, pp. 227--245, 2021.

\bibitem{lee2014resource}
D.~H. Lee, S.~A. Zaheer, and J.~H. Kim, ``A resource-oriented, decentralized
  auction algorithm for multirobot task allocation,'' \emph{IEEE Transactions
  on Automation Science and Engineering}, vol.~12, no.~4, pp. 1469--1481, 2014.

\bibitem{ha2018min}
Q.~M. Ha, Y.~Deville, Q.~D. Pham, and M.~H. H{\`a}, ``On the min-cost traveling
  salesman problem with drone,'' \emph{Transportation Research Part C: Emerging
  Technologies}, vol.~86, pp. 597--621, 2018.

\bibitem{moore2007distributed}
B.~J. Moore and K.~M. Passino, ``Distributed task assignment for mobile
  agents,'' \emph{IEEE Transactions on automatic control}, vol.~52, no.~4, pp.
  749--753, 2007.

\bibitem{roberti2021exact}
R.~Roberti and M.~Ruthmair, ``Exact methods for the traveling salesman problem
  with drone,'' \emph{Transportation Science}, vol.~55, no.~2, pp. 315--335,
  2021.

\bibitem{han2020metaheuristic}
Y.~Han, J.~Li, Z.~Liu, C.~Liu, and J.~Tian, ``Metaheuristic algorithm for
  solving the multi-objective vehicle routing problem with time window and
  drones,'' \emph{International Journal of Advanced Robotic Systems}, vol.~17,
  no.~2, pp. 1--14, 2020.

\bibitem{hong2018range}
I.~Hong, M.~Kuby, and A.~T. Murray, ``A range-restricted recharging station
  coverage model for drone delivery service planning,'' \emph{Transportation
  Research Part C: Emerging Technologies}, vol.~90, pp. 198--212, 2018.

\bibitem{peng2019hybrid}
K.~Peng, J.~Du, F.~Lu, Q.~Sun, Y.~Dong, P.~Zhou, and M.~Hu, ``A hybrid genetic
  algorithm on routing and scheduling for vehicle-assisted multi-drone parcel
  delivery,'' \emph{IEEE Access}, vol.~7, pp. 49\,191--49\,200, 2019.

\bibitem{gao2020learn}
L.~Gao, M.~Chen, Q.~Chen, G.~Luo, N.~Zhu, and Z.~Liu. (2020) Learn to design
  the heuristics for vehicle routing problem. [Online]. Available: \url{arXiv
  preprint arXiv:2002.08539}.

\bibitem{kitjacharoenchai2020two}
P.~Kitjacharoenchai, B.-C. Min, and S.~Lee, ``Two echelon vehicle routing
  problem with drones in last mile delivery,'' \emph{International Journal of
  Production Economics}, vol. 225, p. 107598, 2020.

\bibitem{li2020two}
H.~Li, H.~Wang, J.~Chen, and M.~Bai, ``Two-echelon vehicle routing problem with
  time windows and mobile satellites,'' \emph{Transportation Research Part B:
  Methodological}, vol. 138, pp. 179--201, 2020.

\bibitem{ferrandez2016optimization}
S.~M. Ferrandez, T.~Harbison, T.~Weber, R.~Sturges, and R.~Rich, ``Optimization
  of a truck-drone in tandem delivery network using k-means and genetic
  algorithm,'' \emph{Journal of Industrial Engineering and Management (JIEM)},
  vol.~9, no.~2, pp. 374--388, 2016.

\bibitem{liu2020autonomous}
H.~Liu, X.~Li, M.~Fan, G.~Wu, W.~Pedrycz, and P.~N. Suganthan, ``An autonomous
  path planning method for unmanned aerial vehicle based on a tangent
  intersection and target guidance strategy,'' \emph{IEEE Transactions on
  Intelligent Transportation Systems}, pp. 1--13, 2020.

\bibitem{RN43}
Mikrokopter, ``Technical specifications of mk8-3500 standard,''
  \url{http://www.mikrokopter.de/en/products/nmk8stden/nmk8techdaten/}, 1 2017.

\bibitem{jeong2019truck}
H.~Y. Jeong, B.~D. Song, and S.~Lee, ``Truck-drone hybrid delivery routing:
  Payload-energy dependency and no-fly zones,'' \emph{International Journal of
  Production Economics}, vol. 214, pp. 220--233, 2019.

\bibitem{yoon2018traveling}
J.~J. Yoon, ``The traveling salesman problem with multiple drones: an
  optimization model for last-mile delivery,'' Ph.D. dissertation,
  Massachusetts Institute of Technology, 2018.

\bibitem{RN46}
Y.~Lei, S.~Feng, W.~Hui, and H.~Fei, \emph{Analysis of 30 cases of MATLAB
  intelligent algorithm}, 2nd~ed.\hskip 1em plus 0.5em minus 0.4em\relax 37
  Xueyuan Road, Haidian District, Beijing: Beihang University Press, 8 2015,
  ch.~19, pp. 178--187.

\bibitem{gong2018finding}
W.~Gong, Y.~Wang, Z.~Cai, and L.~Wang, ``Finding multiple roots of nonlinear
  equation systems via a repulsion-based adaptive differential evolution,''
  \emph{IEEE Transactions on Systems, Man, and Cybernetics: Systems}, vol.~50,
  no.~4, pp. 1499--1513, 2018.

\bibitem{ropke2006adaptive}
S.~Ropke and D.~Pisinger, ``An adaptive large neighborhood search heuristic for
  the pickup and delivery problem with time windows,'' \emph{Transportation
  science}, vol.~40, no.~4, pp. 455--472, 2006.

\bibitem{masson2013adaptive}
R.~Masson, F.~Lehu{\'e}d{\'e}, and O.~P{\'e}ton, ``An adaptive large
  neighborhood search for the pickup and delivery problem with transfers,''
  \emph{Transportation Science}, vol.~47, no.~3, pp. 344--355, 2013.

\bibitem{2008Research}
C.~Ren, ``Research on {VRPTW} optimizing based on k-means clustering and {IGA}
  for electronic commerce,'' in \emph{IEEE Conference on Industrial Electronics
  \& Applications}, 2008, pp. 61--66.

\bibitem{bae2007integrated}
S.~T. Bae, H.~S. Hwang, G.-S. Cho, and M.~J. Goan, ``Integrated {GA-VRP} solver
  for multi-depot system,'' \emph{Computers \& Industrial Engineering},
  vol.~53, no.~2, pp. 233--240, 2007.

\bibitem{gendreau2008tabu}
M.~Gendreau, M.~Iori, G.~Laporte, and S.~Martello, ``A tabu search heuristic
  for the vehicle routing problem with two-dimensional loading constraints,''
  \emph{Networks: An International Journal}, vol.~51, no.~1, pp. 4--18, 2008.

\bibitem{wang2019vehicle}
Z.~Wang and J.~B. Sheu, ``Vehicle routing problem with drones,''
  \emph{Transportation Research Part B: methodological}, vol. 122, pp.
  350--364, 2019.

\bibitem{murray2020multiple}
C.~C. Murray and R.~Raj, ``The multiple flying sidekicks traveling salesman
  problem: Parcel delivery with multiple drones,'' \emph{Transportation
  Research Part C: Emerging Technologies}, vol. 110, pp. 368--398, 2020.

\end{thebibliography}
		
	\vspace{-8mm}

	\UseRawInputEncoding

\end{document}